\documentclass[twocolumn]{aastex63}
\usepackage{bm}
\usepackage{multirow}
\usepackage{color}
\usepackage{float}

\usepackage{tabularx}
\makeatother
\makeatletter
\usepackage{comment}
\usepackage{natbib}
\usepackage{amsmath}
\usepackage{booktabs}
\usepackage{array}
\usepackage{hyperref}
\hypersetup{colorlinks=true, citecolor=blue, filecolor=magenta,
 urlcolor=cyan, linkcolor=blue}
 
\shortauthors{Long et al.}

\begin{document}	
\defcitealias{sargent12}{S12}
\defcitealias{dacunha15}{D15}
\defcitealias{sch15}{S15}
\defcitealias{michalowski17}{M17}
\defcitealias{popesso19}{P19}
\defcitealias{bisigello18}{B18}
\defcitealias{dudz20}{D20}

\title{Missing Giants: Predictions on Dust-Obscured Galaxy Stellar Mass Assembly Throughout Cosmic Time}

\author[0000-0002-7530-8857]{Arianna S. Long}\altaffiliation{NASA Hubble Fellow; E-mail: arianna.sage.long@gmail.com}
\affiliation{Department of Astronomy, The University of Texas at Austin, Austin, TX, USA}

\author[0000-0002-0930-6466]{Caitlin M. Casey}
\affiliation{Department of Astronomy, The University of Texas at Austin, Austin, TX, USA}

\author[0000-0003-3021-8564]{Claudia del P. Lagos}
\affiliation{International Centre for Radio Astronomy Research (ICRAR), M468, University of Western Australia, 35 Stirling Hwy, Crawley, WA 6009, Australia}
\affiliation{ARC of Excellence for All Sky Astrophysics in 3 Dimensions (ASTRO 3D)}

\author[0000-0003-3216-7190]{Erini L. Lambrides} \altaffiliation{NPP Fellow}
\affiliation{NASA Goddard Space Flight Center, 8800 Greenbelt Road Greenbelt, MD 20771, USA}

\author[0000-0002-7051-1100]{Jorge A. Zavala}
\affiliation{National Astronomical Observatory of Japan, 2-21-1 Osawa, Mitaka, Tokyo 181-8588, Japan}

\author[0000-0002-6184-9097]{Jaclyn Champagne}
\affiliation{Steward Observatory, University of Arizona, 933 N Cherry Ave, Tucson, AZ 85721, USA}

\author[0000-0003-3881-1397]{Olivia R. Cooper}\altaffiliation{NSF Graduate Fellow}
\affiliation{The University of Texas at Austin, Department of Astronomy, Austin, TX, United States}

\author[0000-0002-3892-0190]{Asantha R. Cooray}
\affiliation{Department of Physics \& Astronomy, University of California, Irvine, 4129 Reines Hall, Irvine, CA 92697, USA}

\begin{abstract}

Due to their extremely dust-obscured nature, much uncertainty still exists surrounding the stellar mass growth and content in dusty, star-forming galaxies (DSFGs) at $z>1$. In this work, we present a numerical model built using empirical data on DSFGs to estimate their stellar mass contributions across the first $\sim$\,10\,Gyr of cosmic time. We generate a dust-obscured stellar mass function that extends beyond the mass limit of star-forming stellar mass functions in the literature, and predict that massive DSFGs constitute as much as $50-100\%$ of all star-forming galaxies with M\,$\ge10^{11}$\,M$_\odot$ at $z>1$. We predict the number density of massive DSFGs and find general agreement with observations, although more data is needed to narrow wide observational uncertainties. We forward model mock massive DSFGs to their quiescent descendants and find remarkable agreement with observations from the literature demonstrating that, to first order, massive DSFGs are a sufficient ancestral population to describe the prevalence of massive quiescent galaxies at $z>1$. We predict that massive DSFGs and their descendants contribute as much as $25-60\%$ to the cosmic stellar mass density during the peak of cosmic star formation, and predict an intense epoch of population growth during the $\sim1$\,Gyr from $z=6$ to 3 during which the majority of the most massive galaxies at high-$z$ grow and then quench. Future studies seeking to understand massive galaxy growth and evolution in the early Universe should strategize synergies with data from the latest observatories (e.g. \textit{JWST} and ALMA) to better include the heavily dust-obscured galaxy population. 

\end{abstract}


\section{Introduction}

Most of what we understand about galaxy evolution in the first six billion years of the Universe is largely based on rest-frame ultraviolet-to-optical studies. This includes, but is not limited to: the evolution of the galaxy stellar mass function \citep[e.g.][]{Peng2010, muzzin13, davidzon17, Thorne2021} used to describe the growth of galaxies over time; the large scale structure of galaxies \citep[e.g.][]{Eisenstein2011, Zehavi2011} that demonstrates how galaxies are distributed throughout space as they form and grow; and the star-forming main sequence (e.g. \citealt{whitaker14} and Table 3 in \citealt{speagle}) establishing that the majority of galaxies form their stars steadily, over long periods of time ($\sim1$\,Gyr), at a rate proportional to their stellar mass. These hereinafter referred to as `UV/OP-based' relationships are often perceived as fundamental truths that are either fed into or used as benchmarks for success in testing our best cosmological models \citep[e.g.][]{Somerville2015, Wechsler2018}. In other words, our primary understanding of galaxy evolution -- from observations to simulations -- is deeply and intricately rooted in UV/optical astronomy, including all of its strengths and limitations.

It is now well known that the rest-frame UV/optical spectrum of the Universe contains only a piece of the larger picture on star formation in the cosmos. Nearly \textit{half} of all cosmic starlight in the Universe is obscured by dust \citep{Driver2008}, which re-radiates at longer infrared wavelengths. This phenomenon is most evident in the cosmic star formation rate density where, out to $z\sim3$, the majority of stellar mass is built in regions obscured by dust and thus primarily detected via infrared (IR) observations \citep[see][ for a review]{madau14}. This means that studies using only UV/optical to chart stellar mass assembly throughout cosmic time are incomplete and biased against significantly dusty systems. 

Over the last two decades, wide-field far-IR and submillimeter surveys ($\lambda > 40$\,$\mu$m) have discovered significant populations of dusty, star-forming galaxies (DSFGs, see \citealt*{casey2014dusty} for a review). Though rare in the local Universe, DSFGs are a thousand times more populous at $z\sim1-3$ \citep[e.g.][]{simpson14, Zavala2021}, dominating cosmic star formation during these times. Their extreme rates of star formation ($\sim10^{2-3}$\,M$_\odot$\,yr$^{-1}$) generate massive reservoirs of dust that obscures the starlight, making some DSFGs nearly invisible to even the deepest UV/optical/near-IR observations, but bright like beacons at far-IR and (sub)millimeter wavelengths \citep[e.g.][]{dacunha15, Williams2019, Manning2022}. 

While some DSFGs are detected in the deepest of UV/optical/near-IR surveys, they are generally left out of large surveys broadly describing galaxy evolution. This usually happens through e.g. stringent requirements for photometric coverage over the rest-frame UV/optical spectrum \citep[e.g. requiring \textit{r, i,} and \textit{z}-band detections,][]{sherman20}, and/or ``redder'' selection bands with observations that are still insufficiently deep to capture the heavily obscured (and redshifted) starlight \citep[e.g. K$_s$-band $<$ 24 mag,][]{lee15}. While often the best available tool for such studies, these requirements are biased against sources with significant attenuation in their UV/optical spectral energy distributions (SEDs); and, if DSFGs are captured, they are often treated as contaminants to the broader goals of the study, to be either removed or ignored as an insignificantly small portion of the broader galaxy population \citep[e.g. $<10\%$ at its peak fraction][]{sargent12,Hwang2021}. Thus, it is likely that several of the fundamental UV/OP-based scaling relations used to describe galaxy evolution and inform our finest models and simulations are lacking a proper accounting of the DSFG population -- a population that drives and dominates stellar mass assembly in the first $\sim6$ billion years of the cosmos. 

Indeed, the astrophysical community has been struggling to properly model and reproduce DSFG populations -- and their measured physical characteristics -- since their discovery \cite[e.g.][]{Granato2000, Baugh2005,hopkins10,hayward2013b,aoyama,mcalpine}. This may have to do with a need for more sophisticated dust radiative transfer treatments, degeneracies in constraining DSFG stellar initial mass functions, complex dust and star geometries, and/or a general lack of constraints on the physical origins / triggering mechanisms of DSFGs (see e.g. \citealt{Hayward2021} for an in depth discussion). Many of the same simulations struggling to properly reproduce DSFG population observations also struggle to produce sufficient populations of massive (M\,$\sim10^{11}$\,M$_\odot$) quiescent galaxies at early times \citep[e.g.][]{Brennan2015, Merlin2019}. It is possible that these mysteries and shortcomings are related because it is also likely that these galaxy sub-populations are related.

Several lines of evidence suggest that the brightest DSFGs constitute the primary ancestral population of the most massive quiescent galaxies in the Universe. This includes comparable comoving number densities \cite[n\,$\sim10^{-5}-10^{-6}$\,Mpc$^{-3}$, e.g.][]{toft14, simpson14}, dark matter halo masses and clustering properties \citep[M\,$\sim10^{12-13}$\,M$_\odot$, e.g.][]{hickox2012, Lim2020}, physical shapes and effective radii \citep[r$_\mathrm{eff}\sim2$\,kpc, e.g.][]{hodge16}, star formation histories \citep[strong and bursty with SFRs\,$\gtrsim10^{2-3}$\,M$_\odot$\,yr$^{-1}$over 50-100\,Myr, e.g.][]{Valentino2020}, and stellar masses \citep[$\sim10^{11}$\,M$_\odot$, e.g.][]{dacunha15, miett17}. Prior studies have also demonstrated a clear relationship between stellar mass and dustiness, where the most massive galaxies are more dust-obscured across cosmic time \citep{vanDokkum2006, pannella09, Wuyts2011, Whitaker2012, whitaker17, deshmukh18}. Recent discoveries of massive, $z>3$ quiescent galaxies demand short, bursty star-formation histories in line with those exhibited by DSFGs \citep[e.g.][]{Glazebrook2017, Schreiber2018}. And, finally, the \textsc{shark} semi-analytical model (which is largely successful in modeling on-sky DSFG populations) suggests that the majority of the most massive galaxies (M\,$>10^{11}$\,M$_\odot$) at $z\approx1$ undergo a (sub)millimeter-bright phase at $z>1$ \citep[e.g.][]{Lagos2020}. Thus, in order for us to understand how the most massive galaxies form and quench during the first half of the cosmic time, it is of the utmost importance that we are dogged in our pursuit towards understanding massive, dust-obscured galaxies and placing them into the wider context of massive galaxy evolution at $z>1$. 

The objectives of this work are twofold. The first is to quantify, to first-order, the significance of missing DSFG populations at the massive end of the star-forming galaxy stellar mass function (SMF). Star-forming SMFs are biased against dusty galaxies as they are determined using UV/optical tracers. Infrared luminosity functions, however, primarily trace dust emission from the star-forming regions of galaxies. Combined, the UV/OP and IR luminosity functions make up the cosmic star formation rate density known today \citep{madau14}. Still missing from the picture, however, is a evolutionary depiction of the stellar mass contributions from dust-obscured star-forming galaxies. Using the IR luminosity function as a nearly-complete consensus of DSFGs out to $z\sim 3-4$, we seek to quantify what the dust-obscured stellar mass function might look like, and posit that it would generate more numerous massive galaxies than previously measured in UV/OP-based SMFs.

The second objective is to test the hypothesis that, despite the heterogeneity of the DSFG population, it is sufficient and complete enough to describe and model the assembly of massive quiescent galaxies in the early Universe. Using a suite of simple assumptions to forward evolve mock populations of DSFGs, we are able to reproduce the observed evolution of massive, passive galaxies -- without needing to invoke any complex assumptions on merger histories, feedback / quenching mechanisms, or even contributions from outside populations \citep[e.g. blue nuggets, ][]{miett17}. 

This paper is organized as follows: in Section \ref{sec:ingredients} we describe the basis assumptions and empirical data used to build the model used in our analysis; in Section \ref{sec:smfcreation} we describe the resulting dust-obscured stellar mass function from our model and how we forward evolve mock galaxies to create their quiescent descendants (Sec. \ref{sec:q_smf}); in Section \ref{sec:inthesky} we present the number and stellar mass density predictions and compare to those in the literature; in Section \ref{sec:otherimplications}, we discuss some additional implications of this model, and provide a summary in Section \ref{sec:summary}. Throughout this work, we adopt a \textit{Planck} cosmology, where $H_0 = 67.7$\,km\,s$^{-1}$\,Mpc$^{-1}$ and $\Omega_\Lambda = 0.692$ \citep{PlanckCollaboration2016}; where relevant, we adopt a Chabrier IMF \citep{chabrier}.

\begin{figure*}[ht]
\begin{center}
\includegraphics[trim=0cm 0cm 0cm 0cm, angle=90, width=1.\textwidth]{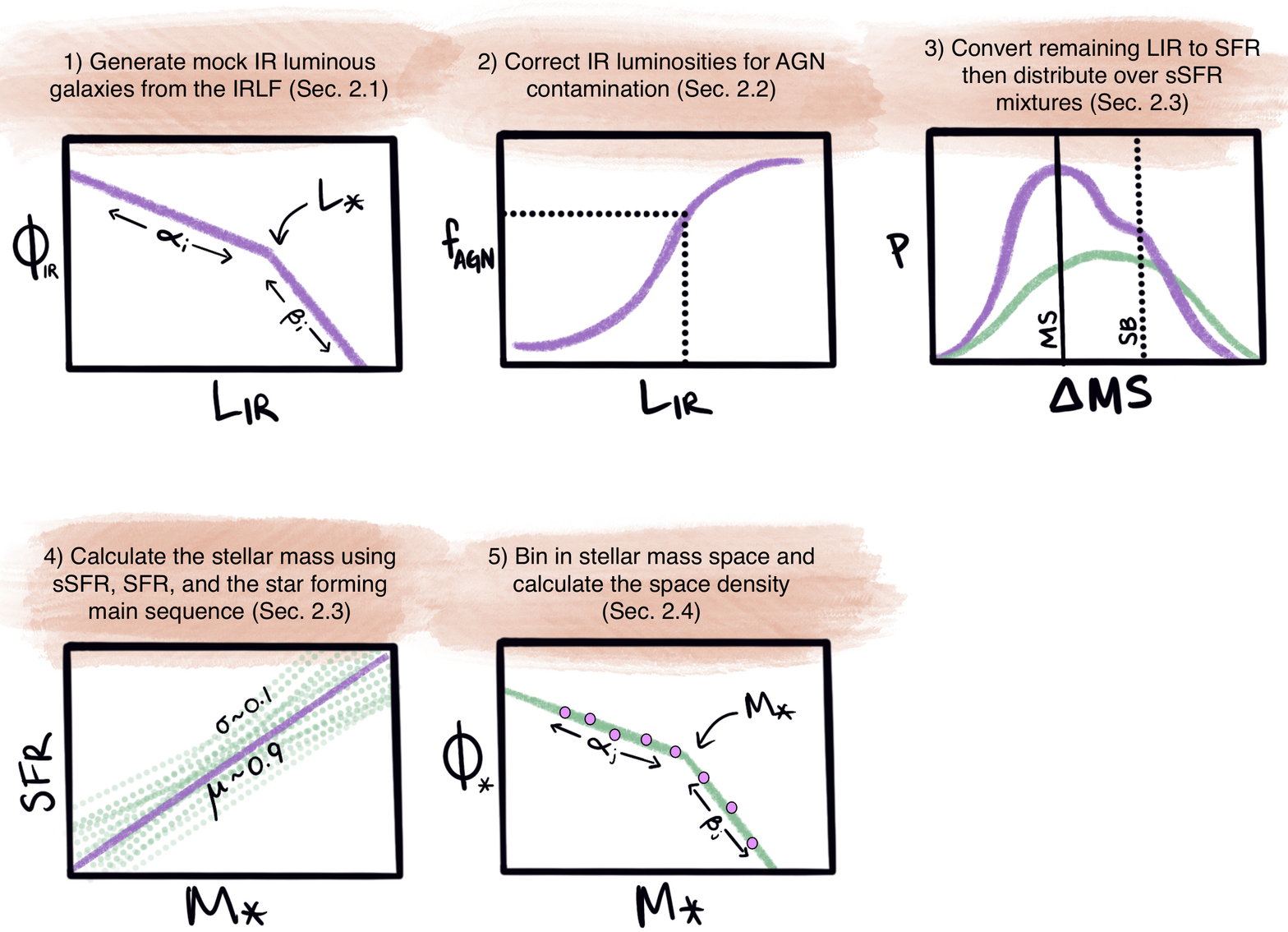}
\end{center}
\caption{A cartoon of the model presented in this work. Several of the parameters highlighted in this cartoon are randomly sampled to account for uncertainties, variation, and unknowns across the literature. \textbf{1)} The model begins by generating mock populations of IR luminous galaxies according the the IR luminosity function defined in \citet{Casey2018b} and \citet{Zavala2021} (Section \ref{sec:IRLF}). \textbf{2)} Then, mock galaxy IR luminosities are corrected to remove AGN contributions in the infrared spectrum (Section \ref{sec:AGN}). \textbf{3)} We then convert the corrected IR luminosities to star formation rates and use these SFRs to randomly distribute galaxies in sSFR space according to various Gaussian mixtures of main-sequence versus starburst galaxies (aka distance from the star forming main-sequence, Section \ref{sec:MvS}; see also Figure \ref{fig:sbfracs}). \textbf{4)} Next, we generate a star forming main-sequence at the coeval redshift of interest (Section \ref{sec:SFMS}) and use this relationship, combined with the SFR and sSFR, to back out the galaxy stellar masses. \textbf{5)} Finally, we re-bin the galaxies in stellar mass space to derive space densities (Section \ref{sec:binning}). This process is repeated a thousand times per 1\,deg$^2$ redshift slice (i.e. $\delta$z) for a given main-sequence vs. starburst ratio, taking care to randomly sample the relevant parameter space (see Table \ref{table:basis}), then combined to derive an overall stellar mass function for dust-obscured galaxies (see Section \ref{sec:DOSMF}). 
} \label{fig:cartoon model}
\end{figure*}

\section{Model Ingredients}\label{sec:ingredients}
Recent theoretical and observational show that the most massive star-forming galaxies in the $z>1$ Universe are heavily obscured by dust. This would mean that a significant portion of cosmic stellar mass is enshrouded in dust, and therefore difficult to detect with traditional photometric methods in the rest-frame UV/optical spectrum. Our goal in this work is to use existing empirical data on dusty, star-forming galaxies to test this hypothesis and quantify the impact of DSFGs on the massive end of galaxy evolution. In this section, we describe the empirical data and assumptions used to build our numerical model, as well as the model itself. There are four main steps to populate the model, each described and motivated in greater detail in the proceeding subsections. We: 
\begin{enumerate}
    \item Use the infrared luminosity function to derive a mock population of DSFGs (Section \ref{sec:IRLF}).
    \item Correct IR luminosities for AGN contamination so that L$_\mathrm{IR}$ accounts for only star formation emission (Section \ref{sec:AGN}).
    \item Convert L$_\mathrm{IR}$ to SFR and distribute SFRs over the star-forming main sequence (i.e. the M$_*$-SFR relation) to derive stellar masses for the mock DSFGs (Section \ref{sec:SFMS}).
    \item Bin in stellar mass space and compute the volume-averaged stellar mass density (Section \ref{sec:binning}). 
\end{enumerate}

In Figure \ref{fig:cartoon model}, we share a cartoon depiction of our model. The model is realized in a light cone of solid angle 1\,deg$^2$ over a given redshift interval; the redshift windows cover $0.5 < z < 6$ and are set to match existing stellar mass functions derived from deep and wide surveys, e.g. \citet{muzzin13} and \citet{davidzon17}. All four steps occur within each of the 10$^3$ realizations, with each realization drawing from probability density functions that mimic uncertainties based on empirical measurements. In Table \ref{table:basis}, we summarize the parameters and their probability distributions that we sample in this model. See the specific subsections that follow for details and motivation on the chosen parameters and distributions. 

Finally, Section \ref{sec:smfcreation} describes how we combine these realizations to create our resulting stellar mass function, and provides the parameters describing the shape of this dust-obscured SMF. 

\begin{deluxetable*}{p{3.cm} p{3.5cm} p{0.5cm} p{9cm}}[ht]
\centering
\tablecaption{Physical relationships and respective parameter spaces used in this model.}
 \tablehead{ 
 \colhead{Physical Relationship} &  \colhead{References} & \colhead{Section} & \colhead{Parameter Space \& Modeled Distribution}}
 \startdata
 Infrared Luminosity Function & \citet{casey18a} \& \citet{Zavala2021} & \ref{sec:IRLF} & Same as in \citet[][Table 3 therein]{casey18a}, except for $\psi(z> z_\mathrm{turn})$ and $\alpha_\mathrm{LF}$, reported in \citet{Zavala2021} as $-6.5^{+0.8}_{-1.8}$ and $-0.42^{+0.02}_{-0.04}$, respectively. We employ skewed Gaussians for both parameters. \\
 AGN Host Selection & \citet{Kartaltepe2010} & \ref{sec:AGN} & f$_\mathrm{AGN\,\,Hosts} = 4-100$\% over an S-shaped curve increasing as a function of L$_\mathrm{IR}$ (defined in Fig. 30, therein). We interpolate over the S-shaped curve in each L$_\mathrm{IR}$ bin. \\
 L$_\mathrm{IR}$ Correction for AGN Hosts & \citet{Kirkpatrick2015} & \ref{sec:AGN} & We sample the CDF of composite and AGN hosts in Fig. 3 therein to randomly assign f$_\mathrm{AGN\,\,MIR}$. Then, we use the quadratic equation (Eqn. 5 therein) to assign f$_\mathrm{AGN\,\,TIR}$. We employ Gaussians for the quadratic equation variables. \\
 Star-Forming Main Sequence & \citet{speagle} & \ref{sec:SFMS} & For the slope, we sample a Gaussian with $\mu=0.9$ and $\sigma=0.1$. We keep the time evolution components (Eqn. 28, therein), employing Gaussians for the parameters related to time. The starburst region is fixed to 0.6\,dex above the main-sequence at a given stellar mass. We fix $\sigma_\mathrm{MS} = \sigma_\mathrm{SB} = 0.3$\,dex. \\
 Main Sequence vs. Starburst Decomposition &
 \citetalias{sargent12,dacunha15, sch15, michalowski17, popesso19, bisigello18, dudz20} & \ref{sec:MvS} & We step through six double Gaussian distributions, each with one mean centered on the main sequence and the other centered on the starburst regime ($+0.6$\,dex). The distributions match the following starburst fractions: 50, 40, 30, 20, 10, and 0\%. \\ 
\enddata 
\end{deluxetable*} \label{table:basis}

\subsection{The Infrared Luminosity Function}\label{sec:IRLF}

We seed our model by generating mock populations of IR luminous galaxies according to the IR luminosity function (IRLF), with the structure described by \citet{Casey2018b, casey18a} and updated parameters measured in \citet{Zavala2021}. This IRLF is empirically-constrained at $z<3$ by well-known DSFG spectral energy distributions and physical properties derived using \textit{Spitzer} and \textit{Herschel} data \citep{casey18a}, and at $z>3$ using ALMA observations \citep{Zavala2021}. Below, we briefly describe the IRLF parameterization and constraints used in this work, and refer the reader to \citet{casey18a, Casey2018b} for more details. 

The IRLF in this model takes the form of a double power-law such that:

\begin{align*}
    \Phi(L,z) = \begin{cases}
    \Phi_\star(z)\Big(\frac{L}{L_\star(z)}\Big)^{\alpha_{\mathrm{LF}}(z)} & \mathrm{if}\,\,L < L_\star  \\
    \Phi_\star(z)\Big(\frac{L}{L_\star(z)}\Big)^{\beta_{\mathrm{LF}}(z)} & \mathrm{if}\,\,L \ge L_\star  
    \end{cases} \stepcounter{equation}\tag{\theequation}\label{eqn:irlf}
\end{align*}

\noindent where $L_\star$ (in L$_\odot$) represents the evolving characteristic `knee' of the luminosity function, $\alpha_{\mathrm{LF}}$ represents the \textit{faint-end} slope of the IRLF (below $L_\star$), $\beta_\mathrm{LF}$ represents the \textit{bright-end} slope of the IRLF (above $L_\star$), and $\Phi_\star$ (in Mpc$^{-3}$ dex$^{-1}$) represents the evolving characteristic number density of the luminosity function. 

Both L$_\star$ and $\Phi_\star$ evolve with redshift, reaching a ``turning point'' at $z_\mathrm{turn} \sim 2$ (corresponding to the peak of the CSFRD) such that, at $z > z_\mathrm{turn}$, L$_\star$ and $\Phi_\star$ evolve with a different slope than at $z < z_\mathrm{turn}$. In other words, L$_\star$ and $\Phi_\star$ evolve as $(1+z)^\gamma$ and $(1+z)^\psi$, respectively, with $\gamma$ and $\psi$ changing values at $z \approx z_\mathrm{turn}$. 

In general, the parameters describing the IRLF at $z \lesssim z_\mathrm{turn}$ are observationally well-constrained. We use the values listed in Table 3 of \citet{casey18a}, with the exception of $\psi$ at $z>2$ for which we use the value measured in \citet{Zavala2021}, and refer the reader to these works for more details on how these values were determined. Specifically, we fix the bright-end slope to $\beta_\mathrm{LF} = -3.0$, the redshift evolution of L$_\star$ to $\gamma = 2.8$, and the redshift evolution of $\Phi_\star$ to $\psi = 0.0$. Together, these values successfully reproduce the measured IRLF, L$_\star$, $\Phi_\star$, and the IR luminous galaxy contributions to the CSFRD at $0<z\lesssim2$ \citep[see Appendix A.1 in][]{casey18a}. 


At $z > z_\mathrm{turn}$, there is much less information on the evolution of L$_\star$ and $\Phi_\star$. For L$_\star$, studies report $\gamma$ values as steep as $\sim1.6$ \citep{Gruppioni2013} and as flat as $0.2$ \citep{Koprowski2017}; this IRLF assumes a redshift evolution of $\gamma = 1$, which is in line with the assumption that L$_\star$ evolves towards brighter luminosities at higher redshifts -- as seen in several observational studies \citep[e.g.][]{Gruppioni2013, Koprowski2017, Lim2020}. Higher values (e.g. $\gamma > 1.5$) would create unphysically bright IR luminous galaxies at $z>4$ that are not known to exist. 

For the evolution of $\Phi_\star$ beyond $z_\mathrm{turn}$, studies report $\psi$ values from $\lesssim -10$ \citep{Koprowski2017} to $-0.5$ \citep{Lim2020}, where more positive values correspond to a dust-\textit{rich} early Universe, and more negative values correspond to a dust-\textit{poor} early Universe. In other words, a dust-poor early Universe would manifest as a CSFRD that is \textit{not} dominated by obscured star formation at $z > z_\mathrm{turn}$, and there would be significantly fewer $z>3$ DSFGs in blank-field sub-mm surveys than there are at $z\sim2$. In this work, we adopt the value presented in \citet{Zavala2021} where $\psi = -6.5^{+0.8}_{-1.8}$, corresponding to a more dust-poor early Universe. This $\psi$ value reproduces the most recent ALMA number counts at 1, 2, and 3\,mm, as well as number counts from single dish telescopes at 700\,$\mu$m\,$-$\,1.1\,mm.

The average faint-end slope collated in \citet{casey18a} is $\alpha_{\mathrm{LF}}=-0.6$. However, recent works suggest a potentially shallower slope of $\alpha_{\mathrm{LF}} = [-0.2, -0.5]$ \citep{zavala18, Koprowski2017, Gruppioni2020, Lim2020, Zavala2021}. For this work, we adopt the value reported in \citet{Zavala2021} where $\alpha_{\mathrm{LF}} = -0.42^{+0.02}_{-0.04}$; this value is aligned with the best fit faint-end slopes in \citet{Koprowski2017}, \citet{ zavala18}, and \citet{Lim2020}. 

For this analysis, we randomly sample the parameter space for $\psi$ and $\alpha_{\mathrm{LF}}$ as listed above and defined in \citet{Zavala2021}. The parameters are modeled as skewed Gaussians, with values drawn for every realization.

In review, for each of the 10$^3$ model realizations (i.e. 10$^3$ lightcones), we randomly sample IRLF parameter values from their aforementioned distributions, then use the IRLF to generate a mock IR luminous galaxy population. To determine the number of IR luminous galaxies in a light cone slice ($n_\mathrm{L_{IR}}$), we integrate the IRLF over bins of width $\delta$L$_\mathrm{IR} = 0.25$\,dex down to an IR luminosity of $10^{9}$\,L$_\odot$. To assign IR luminosities in a way that accurately reflects the shape of the IRLF, we then use the IRLF to generate a cumulative distribution function (CDF) for the respective L$_\mathrm{IR}$ bin. Then, we draw $n_\mathrm{\delta L_{IR}}$ galaxies from a random uniform distribution applied to the CDF ordinate. We use the randomly generated IR luminosities to calculate corresponding star formation rates according to Equation 12 in \citet[][originally derived in \citealt{Murphy2011}]{kennevans12} using the variables listed in Table 1 therein for a total IR (TIR) to SFR conversion. We note that not \textit{all} IR luminosity in a DSFG can be attributed to star-forming processes; some likely hails from AGN accretion activity. In the following section, we detail how we model potential AGN hosts and adjust their IR luminosities accordingly. 

\subsection{AGN Contributions to Total Infrared Luminosity}\label{sec:AGN}

The integrated IR luminosity of galaxies encapsulates dust emission heated by both star formation activity \textit{and} active black hole growth (i.e. active galactic nuclei, or AGN). Young stars embedded in dusty nebulae emit primarily UV and optical light that's absorbed by the enshrouding dust. This warmed dust re-emits in the infrared spectrum. Similarly, actively growing coeval black holes are often surrounded by obscuring clumps of dust and gas that's heated by high energy X-ray, UV, and optical photons, and also re-emits in the infrared. Theoretical and observational studies show that the bright end of the IRLF (L$_\mathrm{IR} > 0.1-1 \times 10^{13}$\,L$_\odot$, depending on the epoch) is likely dominated by AGN-powered IR emission \citep{hopkins10,Fu2010,Gruppioni2013,Symeonidis2018, Symeonidis2021}, though AGN-driven IR emission is probably only significant at rest-frame $\lambda = 8-30\,\mu$m wavelengths \citep{itme}. Similarly, studies on samples of IR luminous galaxies demonstrate that the higher the L$_\mathrm{IR}$, the more likely an AGN is present \citep{Kartaltepe2010, Juneau2013, Lemaux2014} -- with an occurrence rate of $>50\%$ at L$_\mathrm{IR} \gtrsim 10^{12}$\,L$_\odot$ at $0<z<4$. Together, these lines of evidence indicate that we cannot assume that all IR emission from our mock galaxies hails only from star formation processes. Since we wish to use IR luminosity as an indicator for star-formation activity, we must correct for AGN contamination in the total infrared spectrum ($8-1000$\,$\mu$m). 

To create a realistic correction effect, we must first define what fraction of IR luminous galaxies likely host an active supermassive black hole (regardless of how IR luminous the AGN is). For this step, we use the AGN population fraction in 70$\mu$m bright sources as a function of IR luminosity reported in \citep[][Fig. 30 therein]{Kartaltepe2010}. This analysis included AGN selected through a myriad of techniques -- X-ray detections, radio loudness, IRAC colors, and mid-IR power laws. At IR luminosities $<10^{12}$\,L$_\odot$, less than 20\% of the IR luminous galaxies host significantly bright AGN; but between $10^{12-14}$\,L$_\odot$, the occurrence rate of AGN jumps from 60\% to nearly 100\%. We note that other, more recent studies on AGN-SF host galaxy fractions report \textit{smaller} AGN occurrence rates as a function of IR luminosity \citep[e.g.][]{Lemaux2014, Symeonidis2021}. Since we're using total IR luminosity as an indicator for star-formation, which in turn will be used to later derive stellar masses (see Section \ref{sec:SFMS}), smaller AGN occurrence rates will result in larger stellar masses in this model. Thus, by using the larger AGN population fractions in \citet{Kartaltepe2010}, we are choosing a more conservative approach that is less likely to predict an unphysical overabundance of massive star-forming galaxies. 

After selecting a subset of mock IR luminous galaxies to host AGN, we must then decide exactly how much of the IR radiation comes from the AGN. For this step, we use the detailed analysis reporting mid and total IR contributions by AGN in ULIRGs at $z=0.3-2.8$ by \citet{Kirkpatrick2015}. We use the distribution of mid-IR fractions (determined by \textit{Spitzer} IRS mid-IR spectral decomposition, Fig. 3 therein) for designated composite and AGN-dominated galaxies as a probability distribution, and draw from this probability distribution to assign mid-IR AGN fractions to AGN hosts. Then, we use the quadratic relationship between mid-IR AGN fractions and total-IR AGN fractions to determine what percent of the total IR emission can be assigned to star formation (Equation 5 and Fig 14, therein). The majority ($\sim70\%$) of AGN hosts in \citet{Kirkpatrick2015} exhibit mid-IR spectral features indicative of AGN mid-IR dominance (where f$_\mathrm{AGN, mid-IR} \ge 60\%$); it is only at this mid-IR fraction that AGN contamination to the \textit{total} IR luminosity ($8-1000$\,$\mu$m) becomes significant ($\ge20\%$). 

In general, this procedure has the most significant impact on our most luminous galaxies of L$_\mathrm{IR} \ge 10^{13}$\,L$_\odot$ -- which often correspond to the most massive galaxies -- with an average AGN total IR fraction of $\sim50$\%. Note that the \citet{Kirkpatrick2015} relationship between mid-IR and total IR AGN dominance maxes out at f$_\mathrm{AGN,\,mid-IR} = 100\%$, corresponding to f$_\mathrm{AGN,\,total IR} \approx 60\%$. Some studies argue that AGN can be powerful enough to dominate the \textit{entire} IR spectrum \citep[e.g.][]{Tsai2015, symeo16,Symeonidis2018} while other studies argue AGN contributions are predominantly limited to the mid-IR \citep[e.g.][]{itme}. Additionally, recent results demonstrate the existence of a population of moderate to powerful AGN which were previously missing from these major AGN-galaxy co-evolution studies as they were assumed to be weak AGN but are instead likely heavily obscured \citep{Lambrides2020}. Together, these studies indicate that the true underlying relationship between galaxy IR luminosities and AGN processes is still unknown. 

In summation, we use a suite of simple assumptions from the combined works of \citet{Kartaltepe2010} and \citet{Kirkpatrick2015} to correct a modest fraction of IR luminosities under the assumption that the IR emission is significantly powered by an AGN. These corrected IR luminosities are then used to calculate star formation rates (as detailed at the end of Section \ref{sec:IRLF}). In future work, we will explore more detailed treatments and characterization of AGN in massive dusty galaxies -- and the consequences of these treatments and assumptions -- in the context of this model.

\subsection{The Star-Forming Main Sequence}\label{sec:SFMS}

One of the most ubiquitous and well-studied relationships describing galaxy evolution is the correlation between a galaxy's star formation rate and stellar mass. This tight relationship, known as the star-forming main sequence (SFMS), suggests that most star-forming galaxies sustain steady (i.e. non-bursty) SFRs over several billion years \citep{Noeske2007a,Noeske2007b, daddi07a}, at the end of which a galaxy exhausts its gas supply and moves off the main sequence towards quiescence. The correlation exists unambiguously across several orders of magnitude in stellar mass throughout cosmic time -- from $z\sim0$ \citep[e.g.][]{Leroy2019, Popesso2019a} to as high as $z\sim4$ \citep[e.g.][]{sch17, santini17, pearson18, Thorne2021}, theoretically making the SFMS useful in predicting stellar masses for a population galaxies with known SFRs, or vice versa. 

The exact shape of the SF main sequence and its evolution through time is still the subject of intense debate. There is general agreement that the MS takes the form of a power-law (i.e. SFR\,$\propto$\,M$_*^\alpha$) with a tight intrinsic scatter of $0.2-0.35$\,dex at M$_* \lesssim 10^{10}$\,M$_\odot$ \citep[e.g.][]{speagle, sch15, Matthee2019}. In this same stellar mass regime, recent studies report that the SFMS slope, $\alpha$, lies between $0.8-1$ \citep[e.g.][]{speagle, lee15, sch15, tomczak16}. However, there is less consensus for galaxies with stellar masses $>10^{10}$\,M$_\odot$ -- which is the range of stellar mass we are primarily interested in for this work.

Several studies describe a flattening of the main sequence above a certain ``turnover'' mass, typically reported at $\sim$\,$3\times10^{10}$\,M$_\odot$ with possible evolution towards higher values during earlier times \citep[e.g.][]{whitaker14, tasca15, sch15, lee15, tomczak16}. Above this mass, the main sequence is typically parameterized as a power law with a slope that is more shallow than the slope in the low-mass regime, effectively saturating in SFRs at high stellar mass. This bending could be driven by physical phenomena such as star formation suppression at high-masses or increased bulge contributions \citep[e.g.][]{whitaker15, Pan2017, Leslie2020, Thorne2021}, and/or by observational effects such as contamination by passive galaxies \citep[e.g.][]{Johnston2015}, biased/incomplete SFR indicators \citep[e.g.][]{dunlop17}, or biased/incomplete SF galaxy sample selection techniques \citep[e.g.][]{Rodighiero2014}. 

For this work, we assume that the turnover at high masses is driven by observational effects, and that the true SFMS is a singular power law. The majority of studies reporting a high mass turnover selected their star-forming galaxy samples using optical / near-IR color cuts \citep{whitaker14, sch15, lee15, tomczak16}, such as the \textit{UVJ} technique \citep{wuyts07, Whitaker2011}. However, color diagnostics centered in the ultraviolet to near-IR regime, while often the best available tool for such studies, are imperfect identifiers of dust obscured star-forming galaxies. Depending on colors and sample characteristics, anywhere from 5-50\% of M$_\star > 10^{10}$ dust obscured galaxies could be misclassified as quiescent due to their dust-reddened spectra \citep{sch15, DominguezSanchez2016, Patil2019, Santini2019, popesso19, Hwang2021}. This misclassification effectively removes some of the most dust obscured, and therefore star-forming, massive galaxies from the analysis, whose SFRs would likely increase the high-mass SFMS slope. Furthermore, the high mass slope can also be analytically flattened by the misclassification of passive galaxies as star-forming \citep{Johnston2015}. 

We note that if we do include a turnover component, extremely IR luminous galaxies in this model will scatter to unphysically high stellar masses ($>10^{13}$\,M$_\odot$). As described later in this section, we use IR-derived SFRs to pivot over the SF main sequence to assign stellar mass. A more IR luminous galaxy translates to a more star forming galaxy, which translates (generally) to a more massive galaxy. However, with the turnover, which is typically placed around M$_\star \sim 10^{10.5}$\,M$_\odot$ and essentially flattens out from there, most galaxies at and beyond ULIRG luminosities (L$_\mathrm{IR} > 10^{12}$\,L$_\odot$) have SFRs beyond the SFR that corresponds to the turnover mass. In other words, the SFRs exhibited by some of the most extreme DSFGs are well beyond the turnover SFR that corresponds to the turnover mass and therefore have no ``anchor'' to pivot over. Thus, in this model, employing a turnover would require forcing highly star forming DSFGs to scatter around the asymptotic turn over and thereby generate an unphysically abundant population of incredibly massive galaxies. 


Studies that sought specifically to constrain the stellar properties of DSFGs find these galaxies sitting on or above a proportionately more massive and more star-forming extension of a singular power law SFMS \citep[e.g.][]{dacunha15, martis16, michalowski17, pearson18}. Recently, a growing body of literature is starting to establish the existence of massive, optically-dark dust obscured galaxy populations that were previously undetected in deep UV/optical/near-IR surveys \citep[e.g.][]{Spitler2014, Wang2019, Casey2019, Williams2019, Toba2020, Smail2021, Manning2022} -- the same surveys used to establish a global SFMS. It is therefore likely that the true underlying SFMS -- one that includes massive dust obscured galaxies -- takes the form of a power law all the way through the high stellar mass regime. 

For this analysis, we randomly sample the power law slope of the SFMS from a Gaussian distribution centered at $\mu_\alpha=0.9$ with a dispersion of $\sigma_\alpha = 0.1$. Using these slopes as opposed to more shallow ones is a more conservative approach; since we pivot SFRs over the main-sequence power law to derive stellar masses, employing shallower slopes (e.g. $\alpha = 0.4-0.6$) generates large, unphysical excesses of massive dust-obscured galaxies. Furthermore, this approach allows us to marginalize our results over the different competing MS slopes established in more recent literature \citep[e.g.][]{sch15, lee15, tomczak16,bisigello18}. We use the evolving normalization ($\beta$) from the \citet{speagle} SFMS (Equation 28, therein); this form is generally consistent across several studies using different populations of galaxies to determine SFMS properties \citep[e.g.][]{Johnston2015,sch15, tomczak16, Kurczynski2016, pearson18, popesso19}. 

\begin{figure}[t]
\begin{center}
\includegraphics[trim=0cm 0cm 0cm 0cm, scale=0.5]{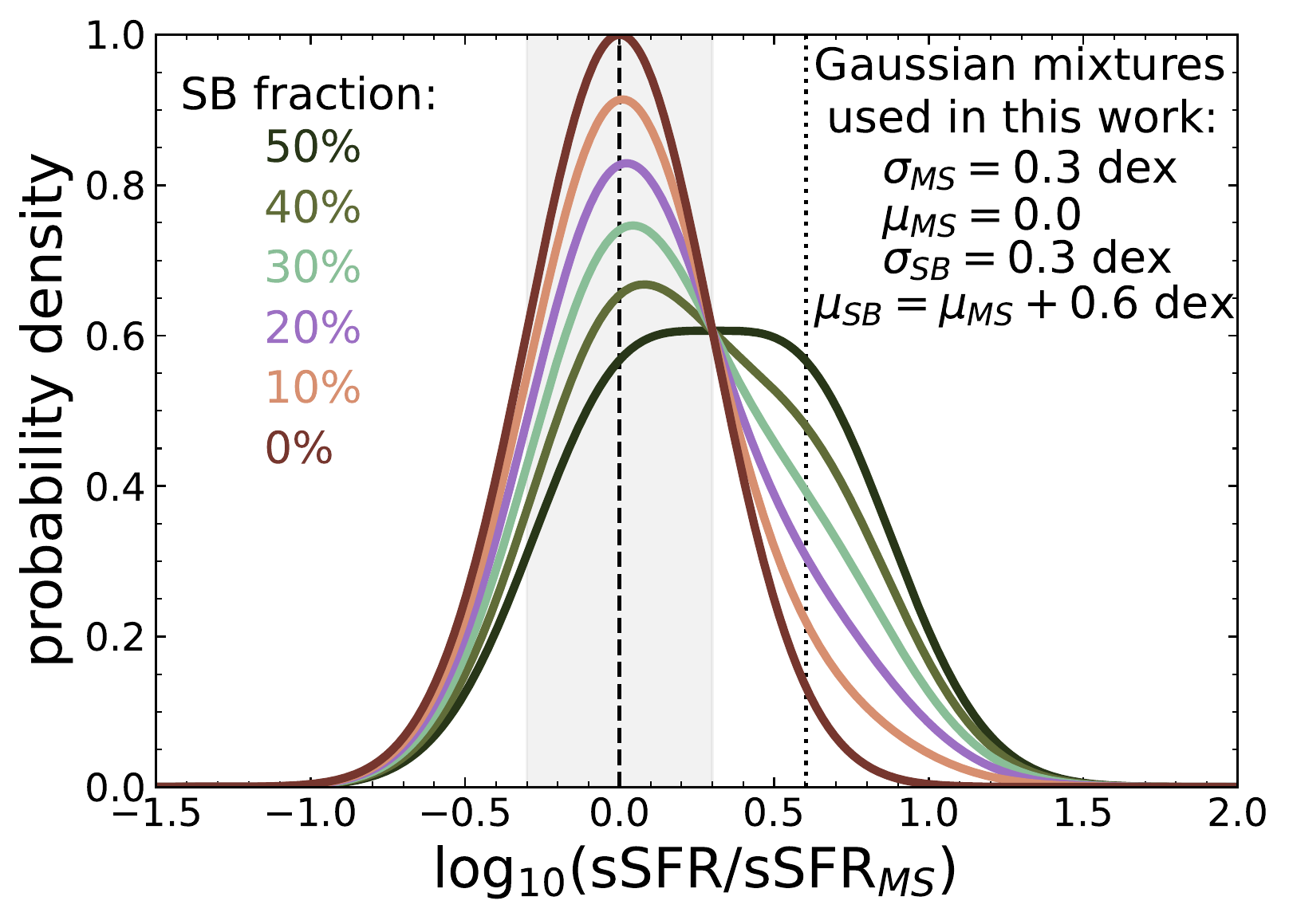}
\end{center}
\caption{The main-sequence and starburst population distributions used to scatter IR luminous galaxies over stellar mass space. Recent studies using more resolved observations with e.g. ALMA \citep[e.g. ][]{dacunha15,michalowski17,dudz20} find that 50-100\% of DSFGs are proportionately more massive and more star-forming extensions of main-sequence galaxies. Since this exact distribution is not well known, we instead step through different Gaussian mixtures that span these observed distributions. For example, the dark green line shows a distribution where half of DSFGs are starbursts and half are main-sequence galaxies, and dark red line shows a distribution where all DSFGs are main-sequence galaxies and none are starbursts. 
} \label{fig:sbfracs}
\end{figure}
\subsubsection{Main Sequence vs. Starburst Decomposition} \label{sec:MvS}

In order to derive stellar masses for our mock IR luminous galaxies, we distribute them over the star-forming main-sequence space. However, there is currently no consensus on exactly where DSFGs as a population lie in main-sequence space. Some studies suggest these objects are primarily starburst galaxies, sitting $\sim$\,4$\times$ above the main-sequence at a given stellar mass, with such extreme SFRs driven potentially by mergers \citep[e.g.][]{Daddi2010, hainline, Lee2013, Magnelli2014}. In this work, we adopt the same starburst galaxy definition.\footnote{The traditional definition of starburst galaxies was based not on their distance from the SFMS, but on their rates of gas consumption (i.e. M$_\mathrm{H_2}$/SFR, also known as the integrated Kennicutt-Schmidt relationship or star formation efficiency, e.g. \citealt{sargent14}). For the purposes of this paper, we use the distance to the main-sequence as the definition.} As a whole, the starburst population is a relatively small portion of the overall star-forming galaxy population \citep[e.g. $\sim3-15\%$ at $z=0-3,$][]{sargent12, sch15, bisigello18, popesso19}, but DSFGs are sometimes believed to be a `starburst-dominated' sub-population of galaxies. Starburst-dominated populations could stem from sensitivity limits by the same surveys, where only the most extreme sources at $z>1$ will be detected. Such a designation is likely also driven by catastrophic projection / blending effects stemming from the large angular resolutions in most canonical wide-field sub-mm surveys built by both space-based \citep[e.g. \textit{Herschel}, ][]{Bethermin2017} and ground-based telescopes \citep[e.g. SCUBA, ][]{cowley15}. 

Recent studies with better angular resolutions (using e.g. ALMA) reveal that significant percentages (50-100\%) of DSFGs are proportionately more massive and star-forming extensions of coeval main-sequence galaxies \citep{dacunha15, michalowski17, Wang2019, dudz20}. This fraction may evolve with redshift, as shown in \citet{dacunha15} and \citet{michalowski17}, but to generalize these fractions as evolutionary trends is problematic as they are dependent on the juxtaposed main-sequence parameterization and the far-IR / sub-mm selection limits; there is also a dearth of sufficiently large samples at epochs beyond $z=2$ to derive a statistically-relevant evolutionary picture. Thus, the true starburst fraction of DSFGs (at least the particularly massive and IR luminous ones) is still unknown, with the truth value likely lying somewhere in between $0-50\%$.

Instead of forcing a singular underlying distribution as truth, we sample the model over several steps within the observed 50-100\% range. Specifically, our model steps through DSFGs-as-starburst fractions of 0, 10, 20, 30, 40, and 50\%, with the remainder belonging to the main-sequence regime. The distribution is modeled as a double Gaussian over main-sequence space, with one Gaussian centered over the main-sequence and the other centered $+0.6$\,dex away \citep[$4\times$ above the main-sequence,][]{rodighiero11}. Both distributions have fixed width of $\sigma_\mathrm{MS} = \sigma_\mathrm{SB} = 0.3$\,dex, similar to those measured in many main-sequence studies \citep[e.g.][]{speagle}. We tested the impact of these chosen widths by running the model with both smaller and larger dispersions ($\sigma = 0.2$ and 0.4, respectively) and found no significant difference ($<3$\%) in our results. These distributions are shown in Figure \ref{fig:sbfracs}. 


Similar to Section \ref{sec:IRLF}, we wish to randomly assign stellar masses in a way that accounts for Eddington bias: if we were to simply draw from the starburst vs. main-sequence Gaussian mixtures as a way to assign stellar mass, more numerous low star-forming galaxies would up-scatter into higher mass bins than high SFR galaxies into lower mass bins. Instead, we model the main-sequence vs. starburst mixtures as cumulative distribution functions, from which we sample a uniform distribution of stellar masses for our mock galaxy populations. This method ensures that galaxies with higher SFRs are more likely to have larger stellar masses, in alignment with observations. 

\subsection{Stellar Mass Space Density} \label{sec:binning}

After using the star-forming main-sequence to generate stellar masses for the mock IR luminous galaxies, we can then bin in stellar mass space to determine the three dimensional stellar mass space density. Galaxies are binned based on their stellar mass in bins of size $\delta_\mathrm{m} = 0.3$\,dex -- corresponding to the average spread in stellar mass in the SFMS \citep{speagle}. The stellar mass histograms are then divided by the same respective light cone volumes used to generate mock galaxies from the IRLF. 

\begin{figure*}[ht]
\begin{center}
\includegraphics[trim=1cm 0cm 0cm 0cm, width=0.33\textwidth]{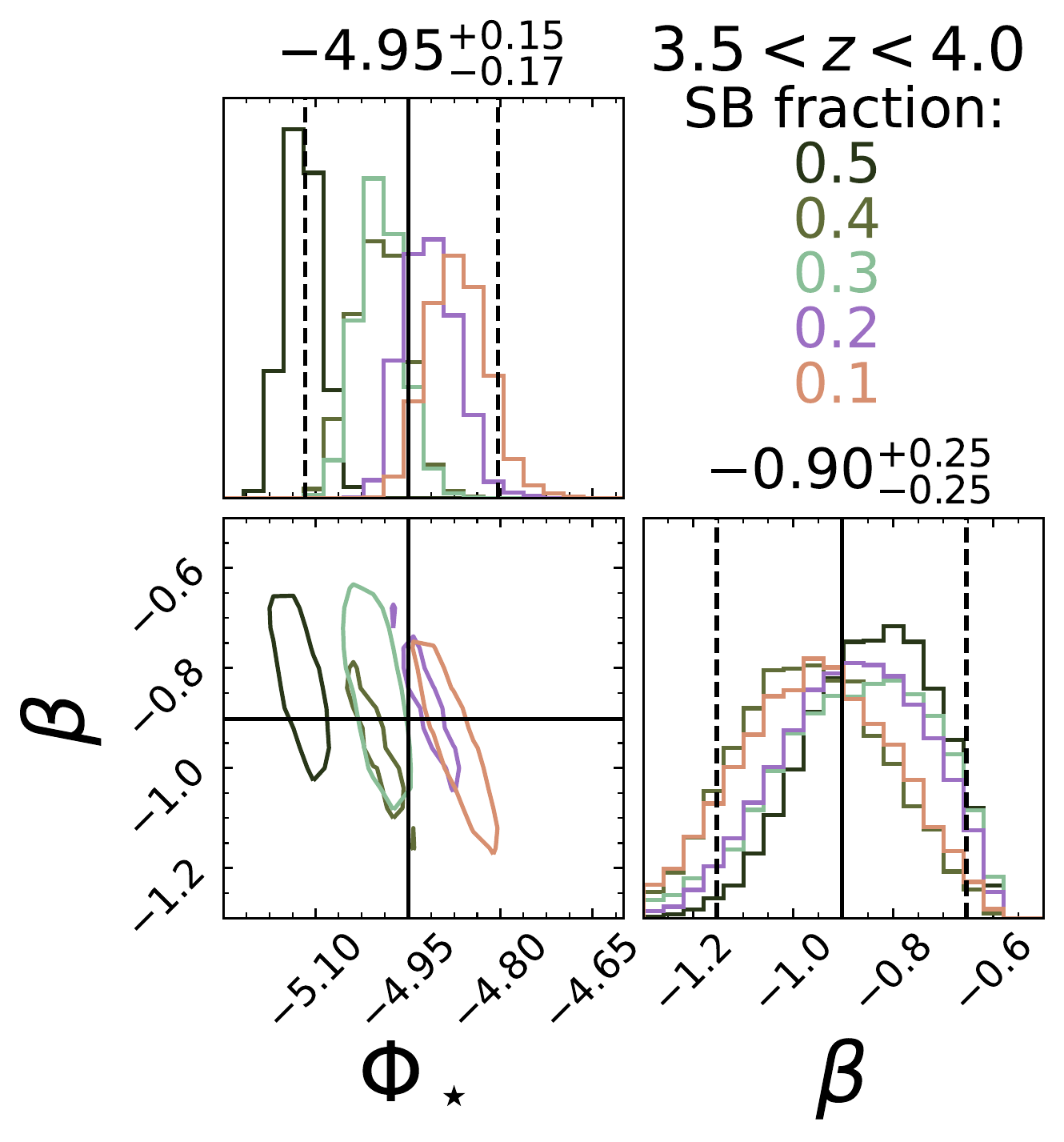}
\includegraphics[trim=0cm 0cm 0cm 0cm, width=0.65\textwidth]{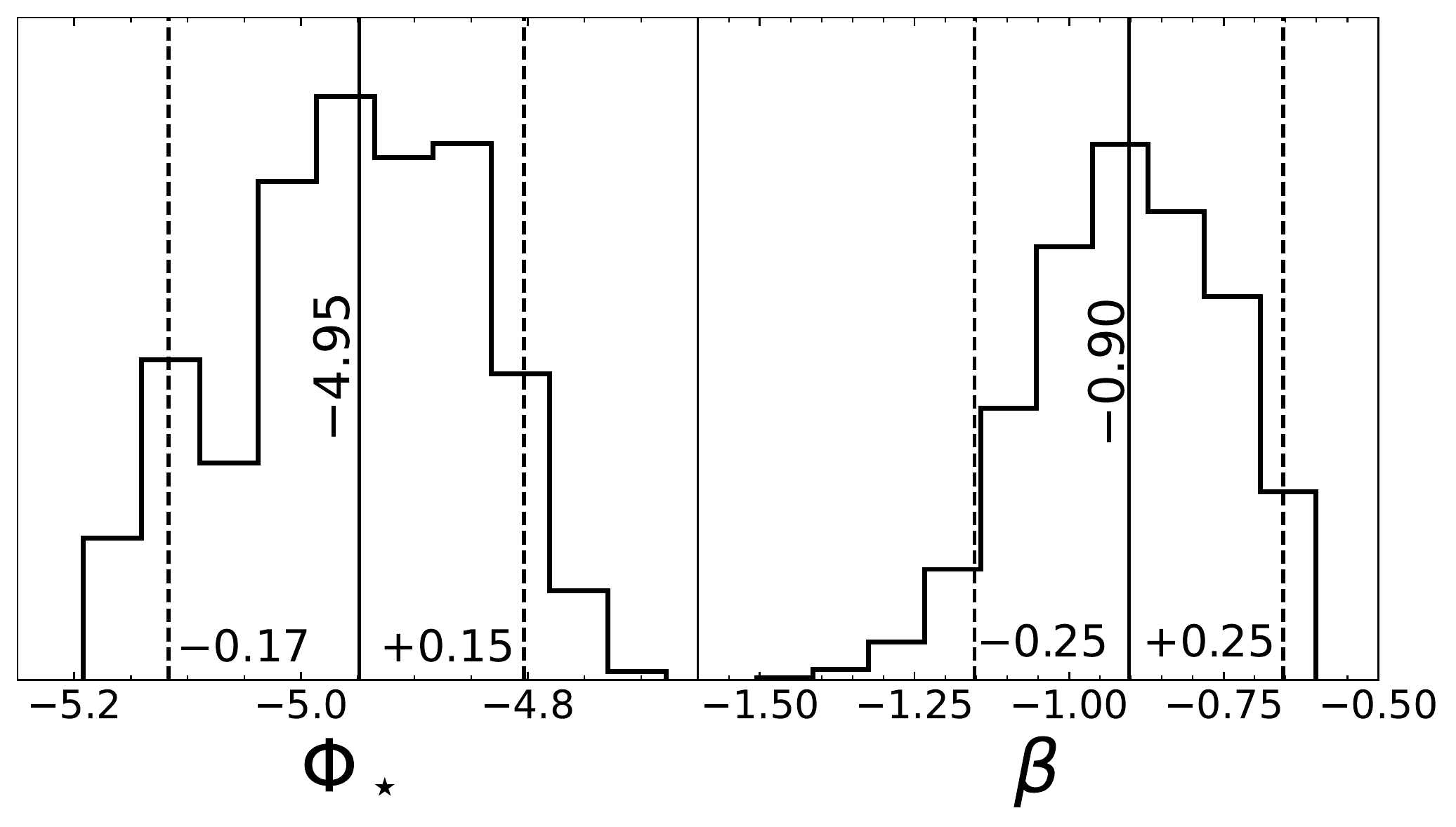}
\end{center}
\caption{\textbf{Left:} Corner (correlation) plots for the stellar mass function parameters ($\Phi_\star$, $\beta$) derived in the $3.5 < z < 4$ redshift bin. Each color represents the best-fit parameters for the six distinct main-sequence vs. starburst mixtures, while the solid line represents the median value and the dash lines represent the 68\% confidence interval (matching the values listed at the top of each histogram). There is some migration of $\Phi_\star$ towards smaller values with increasing starburst fraction, though the differences for both parameters across mixtures are small ($<5\%$) and encapsulated within the 1$\sigma$ uncertainties of the final combined value (shown in the histograms on the right). \textbf{Right:} Combined distributions for $\Phi_\star$ and $\beta$ in the $3.5 < z < 4$ redshift bin. To create these combined distributions, we sampled 10$^3$ successful pairs of $\Phi_\star$ and $\beta$ from each mixture (for a total of $6\times10^3$ samples, a thousand per mixture), then adopted the median value (shown as black solid lines) as the true value for the redshift bin. The black dashed lines show the 1$\sigma$ dispersion \textit{plus} an additional 10\% uncertainty added due to uncertainty in the underlying IRLF; these values are also listed at the top of each parameter column in the corner plot, and at the bottom of the histograms to the right and left of the median value. The ultimate values used in this work are reported in Table \ref{table:bpl}. 
} \label{fig:smfcorner}
\end{figure*}

\begin{figure}[ht]
\includegraphics[trim=0cm 0cm 0cm 0cm, width=0.47\textwidth]{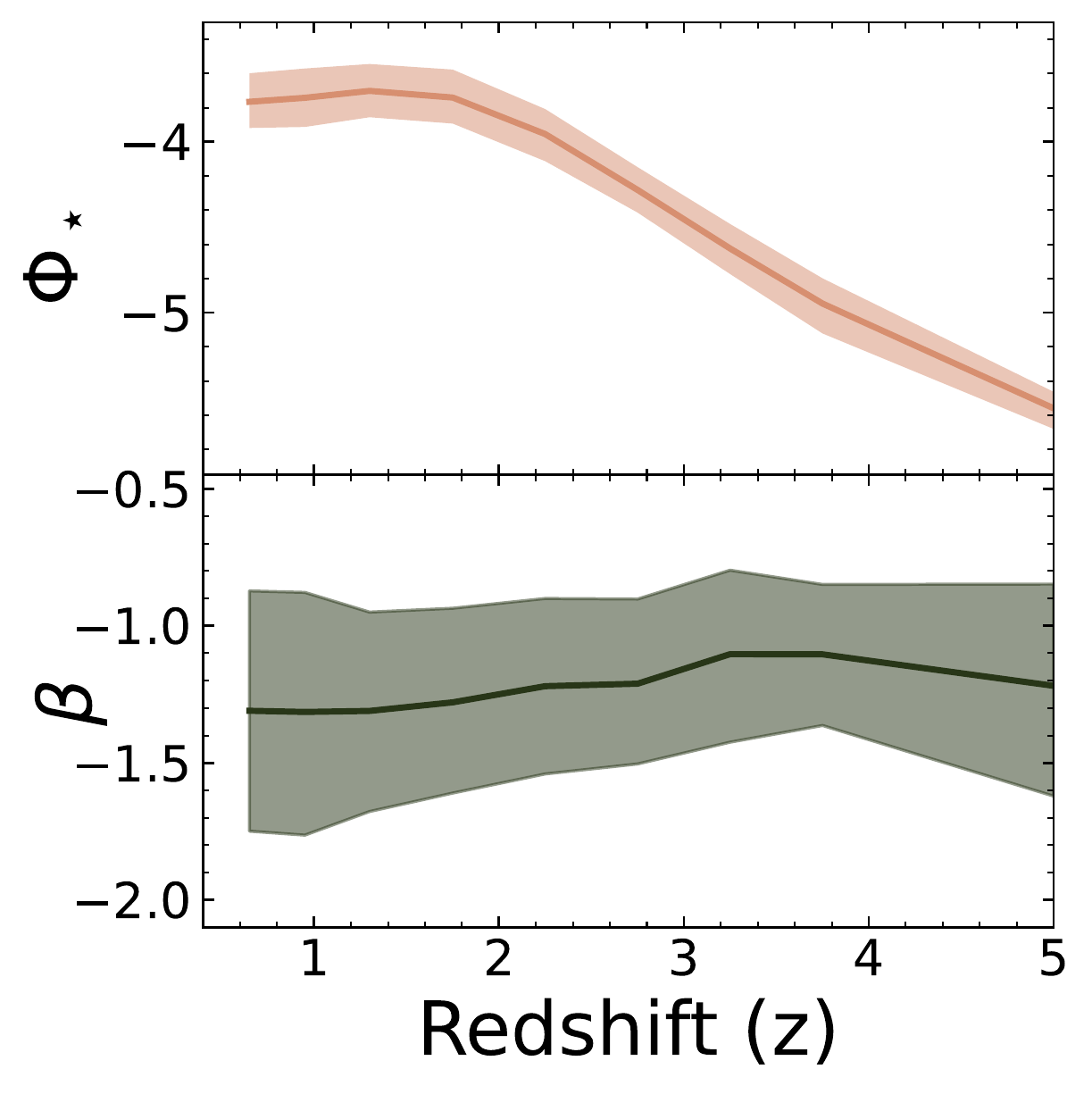}
\caption{The redshift evolution of $\Phi_\star$ and $\beta$ -- the broken power law parameters used to model the DSFG stellar mass function in this work, also listed in Table \ref{table:bpl}. The line marks median value found when combining 10$^3$ pairs of ($\Phi_\star$, $\beta$) from the converged MCMC chains across all starburst vs. main-sequence mixtures, and the shaded area corresponds to the 68\% confidence interval plus an additional 10\% uncertainty. In the top panel, we show the evolution of $\Phi_\star$, the normalization of the space density, decreases at higher-z, which is expected within the context of galaxy growth and assembly. In the bottom panel, we show the evolution of the high mass slope, $\beta$, doesn't evolve significantly with redshift, indicating that massive galaxies grow and then quench out of the massive end of the star-forming SMF over timescales shorter than 1\,Gyr. See Section \ref{sec:dog_smf} for more discussion on these resulting parameters.} \label{fig:smfparams}
\end{figure} 

\section{The Dust-Obscured Stellar Mass Function} \label{sec:DOSMF}

\subsection{Model Fitting Procedure}\label{sec:smfcreation}

To briefly review the model thus far (Sections \ref{sec:IRLF}\,--\,\ref{sec:binning}): we use the IR luminosity function to generate a mock catalog of dust-obscured star-forming galaxies in a given redshift range over 1\,deg$^2$. Redshift bins (listed in Table \ref{table:bpl}) were designed to match SMF studies in the literature for ease in comparison. We correct the IR luminosities for AGN contributions, then convert the corrected IR luminosities to star formation rates. Then, we use the SFRs to pivot the mock galaxies over the star-forming main sequence to generate mock stellar masses. Finally, we bin the mock galaxies in stellar mass space and divide by the light cone volume to calculate stellar mass space densities across cosmic time (as a function of stellar mass). 

This entire procedure is repeated $6\times$10$^3$ times: a thousand realizations for each of the six different main-sequence vs. starburst distributions (described in Section \ref{sec:MvS} and shown in Figure \ref{fig:sbfracs}). Median values of the space densities within each of the six runs are calculated (across the abscissa, i.e. stellar mass bins), and uncertainties determined from the inner 68\% confidence interval. This allows us to directly gauge the impact of varying main-sequence vs. starburst Gaussian mixtures on the resulting stellar mass function. 

Then, within each of the six starburst vs. main-sequence mixtures and for each respective redshift window, we fit a broken power law to the stellar mass space densities. Historically, stellar mass functions are modeled as Schechter functions \citep{schechter76}, but the underlying distribution that our model draws from is a broken power law \citep[the IRLF,][]{casey18a}. Still, we computed the Bayesian information criterion \citep{BIC} to determine whether the Schechter function or a broken power law better models our resulting stellar mass function. The Bayesian information criterion (BIC) is a tool used to choose between two or more models, where models are penalized for the number of parameters included in the fit. A smaller BIC corresponds to the better fit, with a difference of $\delta$\,BIC\,\,$ \ge 10$ indicating a very strong case for the best model. Across all mixtures/redshifts, we found that the broken power law resulted in a stronger model (with $\delta$\,BIC\,$ \ge 7-10$ in nearly all cases). Thus, we adopt the broken power law form to define the number density of dust-obscured galaxies, $\Phi(\mathrm{M})$, as:

\begin{align*}
    \Phi(M,z) = \begin{cases}
    \Phi_\star(z) \Big(\frac{M}{M_\star(z)}\Big)^{\alpha(z)} & \mathrm{if}\,\,M < M_\star(z) \\
    \Phi_\star(z) \Big(\frac{M}{M_\star(z)}\Big)^{\beta(z)} & \mathrm{if}\,\,M \ge M_\star(z)
    \end{cases}
    \stepcounter{equation}\tag{\theequation}\label{eqn:smf}
\end{align*}

\noindent where $\Phi_\star$ (in Mpc$^{-3}$ dex$^{-1}$) is the number density normalization, $\mathrm{M} = \mathrm{M_{star}}/\mathrm{M_\odot}$, $\alpha$ is the low mass slope, M$_\star$ is the characteristic mass or `knee' of the SMF, and $\beta$ is the high mass slope (beyond the knee). We visually inspect the stellar mass space densities in each mixture-redshift combination and find that the low-mass slope ($\alpha$) and the characteristic mass (M$_\star$) do not significantly evolve across mixtures or redshifts ($<5\%$ difference between all mixture-redshift combinations). Therefore, to reduce the degrees of freedom, we fix $\alpha = -0.5$ and M$_\star = 10^{10.8}$\,M$_\odot$ universally across all epochs/mixtures. Upon visual inspection, both of these values combined well-reproduced the low-mass regime of the DSFG stellar mass function for all mixture-redshift combinations. For posterity, we also tested a smaller M$_\star = 10^{10.5}$\,M$_\odot$ and a steeper $\alpha = -0.7$ (more inline with UV/OP-based studies). This resulted in a $<3\%$ difference in the derivations of $\Phi_\star$ and $\beta$, the overall $\chi^2$ values were poorer, indicating preference for a more massive knee and shallower low-mass slope.

Then, still within the individual mixture-redshift combinations, we run 10$^4$ realizations of a Metropolis-Hastings (i.e. MCMC) algorithm to find the best-fit values for the number density normalization, $\Phi_\star$, and massive-end slope, $\beta$. We allow log$_{10}$($\Phi_\star$) to range between $[-6, -3]$ and $\beta$ between $[-2.5, -0.6]$. The quality of each broken power law fit is determined by computing the chi-square goodness of fit against the mock stellar mass space densities; we only compute the $\chi^2$ using stellar mass space densities with 68\% confidence intervals that are less than an order of magnitude away from the median value (i.e. log$_{10}$($\sigma$)\,$ < $\, 1). This primarily affects the highest mass bins (M\,$>10^{12}$\,M$_\odot$) where mock IR luminous galaxies appear rarely and therefore aren't produced consistently in the model. 

In Figure \ref{fig:smfcorner}, we show the resulting constraints on $\Phi_\star$ and $\beta$ for an example redshift bin. There is some marginal covariance with starburst fraction and $\Phi_\star$, such that increasing starburst fractions causes a smaller overall number density of massive IR luminous galaxies (as expected). However, the differences between $\Phi_\star$ across mixtures is small ($\lesssim0.25$\,dex at most) and, moreover, the median $\Phi_\star$ in each mixture is captured within the final uncertainty derived for this redshift bin. Thus, this covariance is insignificant for the purposes of this work. Additionally, there is no clear trend between the high-mass slope $\beta$ and starburst fractions across all epochs. One might expect a shallower $\beta$ to arise from a smaller starburst fraction because with low starburst fractions more galaxies are deemed main-sequence-like and thus scatter to higher stellar masses; this effect in theory creates more numerous massive galaxies than model realizations with ``high'' starburst fractions, and causes the population density near/beyond M$_\star$ to increase. However, in this case, it's likely that the most extreme IR luminous / dust-obscured star-forming galaxies are so rare that a consistently small population emerges at masses $>$\,M$_\star$, regardless of the employed main-sequence vs. starburst mixture.

To derive a singular stellar mass function from the six different main-sequence vs. starburst mixtures, we randomly draw 10$^3$ pairs of ($\Phi_\star$, $\beta$) from the converged MCMC chains of each mixture. We collate the $6\times$10$^3$ pairs of ($\Phi_\star$, $\beta$) and determine the median and 68\% confidence interval. This process is completed separately for each redshift window. We show example collated histograms for $\Phi_\star$ and $\beta$ in the $3.5<z<4$ redshift bin in Figure \ref{fig:smfcorner}. 

It is important to note that the main basis of this model hinges upon the accuracy of the modeled IR luminosity function. As reviewed in Section \ref{sec:IRLF}, the evolution of the IRLF beyond $z\sim2$ is still under debate \citep[e.g.][]{Gruppioni2020}. While we marginalize over several parameters to account for these $z>2$ unknowns, there are still some fixed assumptions in this work, such as the evolution of L$_\star$ at $z>z_\mathrm{turn}$, or the lackthereof for the faint-end slope, $\alpha_\mathrm{LF}$. We find that a steeper $\alpha_\mathrm{LF} = -0.7$ results in a less than $3\%$ difference in the final derivations of the parameters describing the dust-obscured SMF; and a flatter evolution of L$_\star$ with $\psi = 0.5$ results in a less than $1\%$ difference in the resulting SMF parameters. Thus, to be conservative and account for the impacts of our model's main assumptions, we add an additional 10\% uncertainty on the ultimate stellar mass function parameters. This additional uncertainty is included in Figure \ref{fig:smfcorner} and in Table \ref{table:bpl}, where the ultimate SMF parameters for this analysis are listed.

\begin{deluxetable*}{ccccc | cccc}[ht]
\centering
\tablecaption{Parameters for the resulting broken power law functions that model the dust obscured stellar mass function.}
 \tablehead{
 \vspace{-2mm} \\ 
 \multicolumn{1}{c}{} &
 \multicolumn{4}{c}{Dusty Star-Forming Galaxies} & \multicolumn{4}{c}{Quiescent Descendant Galaxies} \\ 
 \colhead{Redshift} & \colhead{log$_{10}(\Phi_{\star}$)} & \colhead{$\alpha$} & \colhead{log$_{10}(\mathrm{M}_\star$)} &  \colhead{$\beta$} & \colhead{log$_{10}(\Phi_{\star}$)} & \colhead{$\alpha$} & \colhead{log$_{10}(\mathrm{M}_\star$)} &  \colhead{$\beta$}}
 \startdata 
$0.5<z<0.8$ & $-3.77$\,$^{+0.17}_{-0.17}$ & $-0.5$ & 10.8 & $-1.30$\,$^{+0.41}_{-0.43}$ & $-3.61$\,$^{+0.16}_{-0.17}$ & $-0.5$ & 10.8 & $-1.44$\,$^{+0.35}_{-0.38}$\\
$0.8<z<1.1$ & $-3.73$\,$^{+0.15}_{-0.19}$ & $-0.5$ & 10.8 & $-1.33$\,$^{+0.43}_{-0.41}$ & $-3.55$\,$^{+0.2}_{-0.19}$ & $-0.5$ & 10.8 & $-1.44$\,$^{+0.39}_{-0.41}$ \\
$1.1<z<1.5$ &  $-3.70$\,$^{+0.16}_{-0.16}$ & $-0.5$ & 10.8 & $-1.31$\,$^{+0.39}_{-0.37}$ & $-3.46$\,$^{+0.16}_{-0.17}$ & $-0.5$ & 10.8 & $-1.42$\,$^{+0.35}_{-0.35}$  \\
$1.5<z<2.0$ & $-3.75$\,$^{+0.17}_{-0.15}$ & $-0.5$ & 10.8 & $-1.28$\,$^{+0.36}_{-0.33}$ & $-3.5$\,$^{+0.18}_{-0.19}$ & $-0.5$ & 10.8 & $-1.37$\,$^{+0.35}_{-0.34}$\\
$2.0<z<2.5$ & $-3.98$\,$^{+0.16}_{-0.14}$ & $-0.5$ & 10.8 &  $-1.20$\,$^{+0.33}_{-0.33}$ & $-3.83$\,$^{+0.19}_{-0.29}$ &  $-0.5$ & 10.8 & $-1.16$\,$^{+0.42}_{-0.3}$\\
$2.5<z<3.0$ & $-4.28$\,$^{+0.13}_{-0.14}$ & $-0.5$ & 10.8 &  $-1.20$\,$^{+0.31}_{-0.3}$ & $-4.11$\,$^{+0.17}_{-0.17}$ & $-0.5$ & 10.8 & $-1.25$\,$^{+0.31}_{-0.27}$  \\
$3.0<z<3.5$ & $-4.63$\,$^{+0.14}_{-0.14}$ & $-0.5$ & 10.8 &  $-1.12$\,$^{+0.32}_{-0.3}$ & $-4.45$\,$^{+0.15}_{-0.15}$ & $-0.5$ & 10.8 & $-1.21$\,$^{+0.28}_{-0.26}$ \\
$3.5<z<4.0$ & $-4.95$\,$^{+0.15}_{-0.17}$ & $-0.5$ & 10.8 &  $-1.10$\,$^{+0.25}_{-0.26}$ & $-4.95$\,$^{+0.13}_{-0.13}$ & $-0.5$ & 10.8 & $-1.09$\,$^{+0.27}_{-0.26}$ \\
$4.0<z<6.0$ &  $-5.56$\,$^{+0.1}_{-0.12}$ & $-0.5$ & 10.8 & $-1.25$\,$^{+0.37}_{-0.40}$ & $-5.43$\,$^{+0.07}_{-0.07}$ &  $-0.5$ & 10.8 & $-0.97$\,$^{+0.18}_{-0.2}$\\
 \enddata
\end{deluxetable*} \label{table:bpl}

\begin{figure*}[ht]
\begin{center}
\includegraphics[trim=0cm 0cm 0cm 0cm, width=1\textwidth]{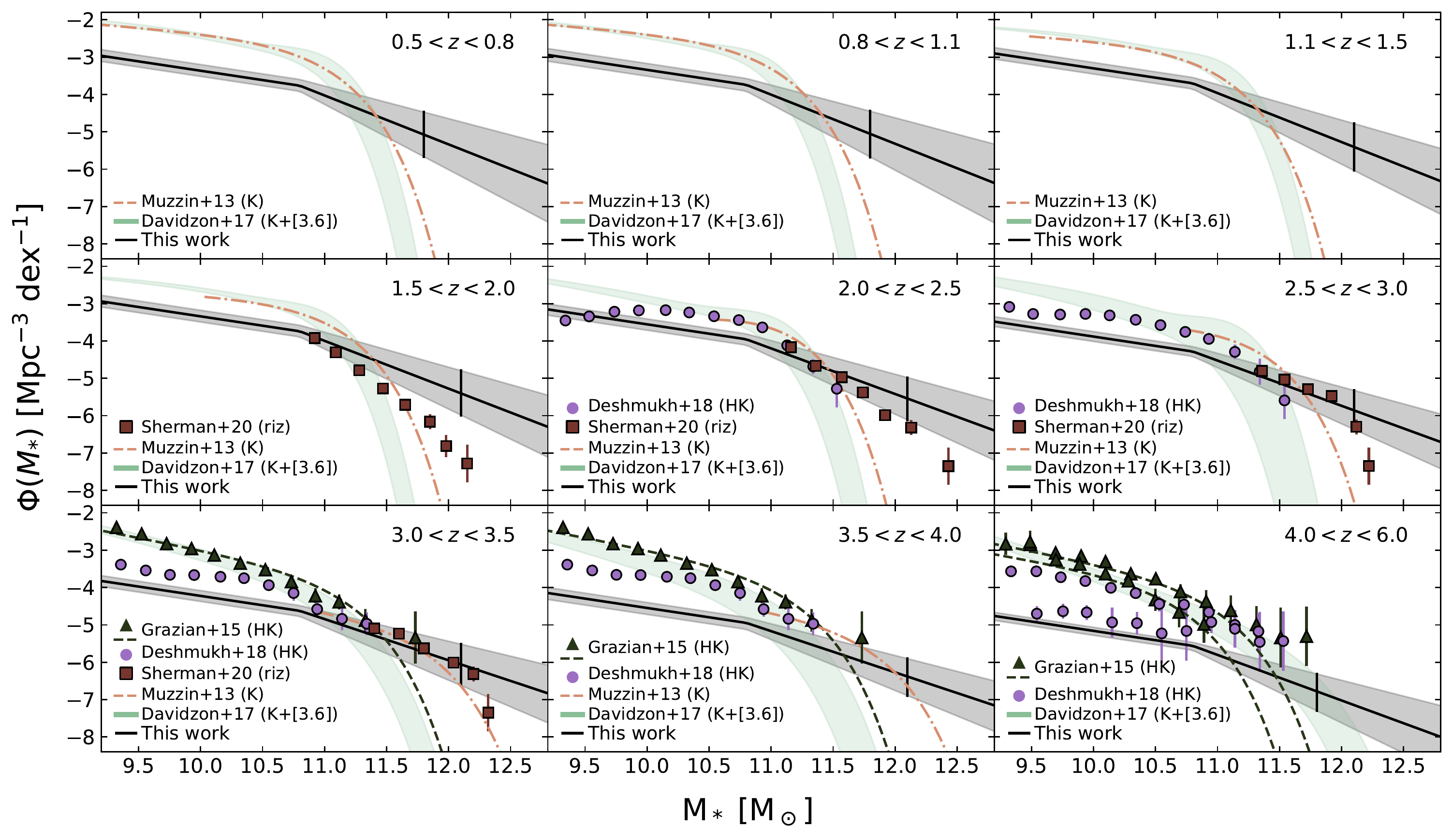}
\end{center}
\caption{The resulting stellar mass function of the IR luminous galaxy population as derived from this model. Our results are shown as the black line, with 1$\sigma$ uncertainty shaded in grey. The vertical black line marks the limit where the uncertainty of the stellar mass space densities is $<$\,1 order of magnitude; beyond this mass, massive IR-luminous galaxies were not consistently produced in the model realizations. Shown for comparison are star-forming SMFs from the literature, with a particular focus on those using the deepest available observational data: \citet[][pink dash-dotted line]{muzzin13}, \citet[][green shaded curve]{davidzon17}, \citet[][red squares]{sherman20}, \citet[][purple circles]{deshmukh18}, \citet[][black triangles and black dashed line]{Grazian2015}. In this work, UV/OP-bright galaxies tend to dominate the star-forming population at lower stellar masses (M\,$\lesssim10^{10.5-11}$\,M$_\odot$) by $\sim$\,an order of magnitude, but become comparatively or less abundant at high stellar masses, particularly at $1<z<4$. See Section \ref{sec:dog_smf} for more details.} \label{fig:smfs}
\end{figure*} 

\subsection{Comparison to Literature SMFs}\label{sec:dog_smf}

In Figure \ref{fig:smfparams}, we show the redshift evolution of $\Phi_\star$ and $\beta$ for our final dust-obscured SMF across time. The number density normalization stays relatively constant at log$_{10}$($\Phi_\star$/(Mpc$^{-3}$\,dex$^{-1}$))\,$ \approx -3.7$ in the 5\,Gyr $0.5<z<2$ redshift window, but experiences nearly an order of magnitude increase in the 1.5\,Gyr between $z=3.5$ and $z=2$ and a similar strength in growth in the 1\,Gyr between $z=6$ and $z=3$. This points to an era of rapid stellar mass growth in the $z>3$ Universe for massive dust-obscured galaxies that is unmatched at later times. Interestingly, however, we do not see any significant evolution in the massive end slope of the SMF, $\beta$, which has an average value of $-1.23\pm 0.08$ across all epochs. Combined, the evolutionary properties for these two parameters paint a picture in line with the cosmic downsizing scenario, where the most massive star-forming galaxies built the majority of their stellar mass in the first 1-2\,Gyr, and quench by $z\sim2$. Furthermore, the rate of growth and quenching in dust-obscured galaxies must occur on short timescales (at least shorter than the $\sim$ 1\,Gyr-wide redshift windows); in other words, if $\beta$ remains relatively stable across time, then dust-obscured galaxies likely grow into and then quench out of the massive end of the star-forming SMF at a nearly equivalent rate. 

In Figure \ref{fig:smfs}, we show the final star-forming stellar mass functions derived from IR luminous galaxies across $0.5<z<6$. We remind the reader that the IRLF does not account for \textit{all} star-forming galaxies and therefore agreement between res-frame UV/OP-based SMFs and the dust-obscured SMF generated in this work is not necessarily expected. Moreover, since our SMF takes a different functional form than those presented in the literature, we cannot quantitatively speak to differences in shape and parameterization. We can, however, discuss qualitative and integrated differences (the latter of which is discussed in detail in Sections \ref{sec:BigDOGnumbers} and beyond). Furthermore, due to the scientific purpose of this work, we focus the majority of this discussion to the more massive end of galaxy evolution at M\,$>3\times10^{10}$\,M$_\odot$, but show the less massive end of this and literature SMFs for the reader to understand the overall extrapolation of this model's results to low masses. 

\begin{figure}[ht]
\begin{center}
\includegraphics[trim=0cm 0cm 0cm 0cm, width=0.45\textwidth]{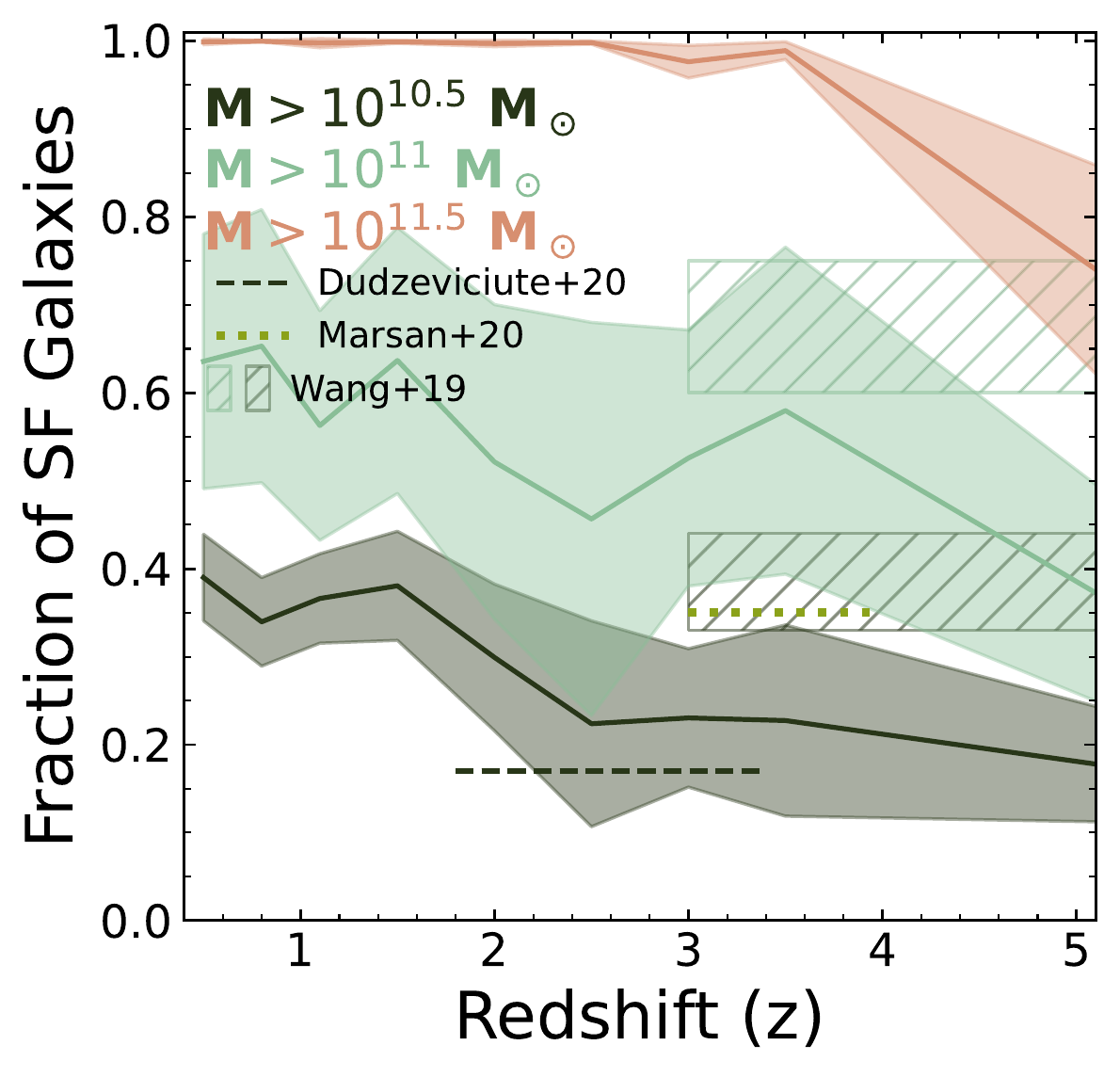}
\end{center}
\caption{Fraction of massive star-forming galaxies that are dust-obscured, defined by LIRG luminosities or greater. The denominator is calculated by adding number counts from the \citet{davidzon17} stellar mass function to the estimates from our dust-obscured SMF. Fractional uncertainties from both SMFs are added in quadrature, with 1$\sigma$ uncertainties shown as the lighter contours. The dark green, light green, and pink bands correspond to lower stellar mass limits of log$_{10}$(M)\,$>10.5$, 11, and 11.5\,M$_\odot$, respectively. Overlaid are comparable results of massive dust-obscured galaxy fractions from \citet{Wang2019}, \citet{dudz20}, and \citet{Marsan2022}. Consistent with the literature, we find that the fraction of dust-obscured SF galaxies increases with increasing stellar mass, reaching totality ($\sim100\%$) at the highest and rarest limit of M\,$\sim10^{11.5}$\,M$_\odot$.
} \label{fig:fractions}
\end{figure}

In general, the dust-obscured SMF produced by this model is consistently below that of UV/OP-based predictions for star-forming galaxies at masses $\lesssim$\,$10^{11}$\,M$_\odot$. It is around this mass that there appears to be a convergence between UV/OP and dusty SMFs, which then turns into a dust-obscured dominated stellar mass function at high masses. As mentioned before, this is likely because low mass galaxies are not accounted for in the IRLF as they are less obscured by dust. The discrepancies between the observed SMFs and our model are greatest at $z=1-4$, where our model posits a significant population of M\,$>10^{11.5-12}$\,M$_\odot$ galaxies while the UV/OP-based estimates see none (see Figure \ref{fig:fractions}). At $z=2-3.5$, our model shows some agreement with that of the $riz$-based SMF derived in \citet{sherman20}. We believe this agreement to be artificial as it could be argued that there are high rates of AGN contamination in the Sherman et al. sample. This is because AGN are removed from this sample using SDSS data which likely does not sufficiently capture high redshift and/or obscured AGN. At $z\ge3$, we also see some partial agreement between our model, the \citet{muzzin13} results, and the massive, dusty galaxy sample within \citet{deshmukh18} out to M\,$\sim10^{11.5-12}$\,M$_\odot$. Both of these works select their galaxy samples using $H$ and/or $K$-band images with observational depths near the limits of known massive, dust-obscured galaxies \citep[e.g. $m_\mathrm{Ks}\approx 23-24.5$][]{simpson14, dacunha15, cowie18, dudz20}, so it is likely some of the massive dust-obscured population is captured in these works. However, at these $H$ and $K$-band depths, only $50-75$\% of dusty, star forming galaxies are typically detected, leaving a significant population of optically faint galaxies  \citep[also known as `OIR-dark',][]{Wang2019, Williams2019, casey19, Manning2022} -- that are likely some of the most massive and heavily dust obscured galaxies in the cosmos \citep{pannella09,dunlop17,whitaker17,fudamoto20} -- unaccounted for in the massive end of these stellar mass functions. 

It is also interesting that our results are in more severe disagreement with the \citealt{davidzon17} analysis as the authors introduce a deep \textit{Spitzer} IRAC [3.6\,$\mu$m] image to help capture more dust-obscured and quiescent galaxies than the typical rest-frame UV/OP-selected SMFs. Yet, the Davidzon et al. SMF yields fewer star-forming galaxies beyond M\,$>10^{11}$\,M$_\odot$ than any other SMF discussed in this section. This could be due to a variety of factors. For example, the Davidzon et al. work required detection in a combined $zYJHK_s$ image; the 3$\sigma$ detection limit of the reddest band (K$_s$ = 24) could miss 40\% or more of dusty galaxies due to their faintness \citep{simpson14, Brisbin2017, dudz20}. Another factor could be the fact that galaxy stellar masses were derived solely from rest-frame UV to optical wavelengths -- no far-infrared (i.e. cold dust) information was included in SED fitting. If there is heavy obscuration of the stellar spectrum, and no energy coupling between optical/mid-IR and FIR observations, stellar masses can be underestimated by a factor of 2$-$3 due to faintness \citep{Buat2019}, and/or a mildly quiescent galaxy spectrum could be assumed (especially with limited photometry). Both of these employed methods can remove massive, dust-obscured galaxies from the massive end of the star-forming SMF.


At $4<z<6$, the dust-obscured SMF is below UV/OP-based estimates at similar redshifts, except where they meet at the most massive point around M\,$\sim10^{12}$\,M$_\odot$. While there is still large uncertainty in the prevalence of dust-obscured galaxies at $z>4$ (see e.g. \citealt{Zavala2021}, c.f. \citealt{Gruppioni2020}), it appears from this model and from UV/OP-based estimates that such massive galaxies are exceedingly rare at these epochs and their population density really begins to build after this epoch. 

In Figure \ref{fig:fractions}, we show the fraction of massive, dust-obscured galaxies among all star-forming galaxies using the \citet{davidzon17} SMF as a proxy for UV/OP-bright star-forming galaxies (i.e. the denominator is derived from Davidzon et al. \textit{plus} the estimate from this work). We show that from $0.5<z<4$ dust-obscured galaxies become more prominent at higher stellar masses. Though there is large uncertainty, we predict that 20-40\% of galaxies at M\,$>10^{10.5}$\,M$_\odot$ are dust obscured, which is similar to the fraction of $K$-band undetected SMGs in \citet{dudz20}, the $H$-band dropout sample in \citet{Wang2019}, and the dust-reddened sample in \citet{Marsan2022}. The percentage increases to 50-60\% at M\,$>10^{11}$\,M$_\odot$, and finally maxes out to nearly 100\% at M\,$>10^{11.5}$\,M$_\odot$. Interestingly, these results are in agreement with the \citet{whitaker17} study, which arrived at a dust-obscured fraction of massive galaxies using a completely different method -- quantifying the fraction of dust-obscured versus UV/OP-bright star formation in individual massive star-forming galaxies at $0<z<2$. At $z>4$, we caution interpretation of the M\,$>10^{11.5}$\,M$_\odot$ DSFG fractions as galaxies of this mass were so rare that they were not consistently produced in this model. 

We note that there may be some massive DSFGs with sufficient escaping stellar emission to be captured in the UV/OP-based estimates and therefore there may be some ``double counting'' of massive SF galaxies in this space. Understanding the significance of that effect is beyond the scope of this paper, and likely unfeasible until sufficient wide-field \textit{JWST} near-IR surveys are completed.

\begin{figure*}[ht]
\begin{center}
\includegraphics[trim=0cm 0cm 0cm 0cm, width=1.\textwidth]{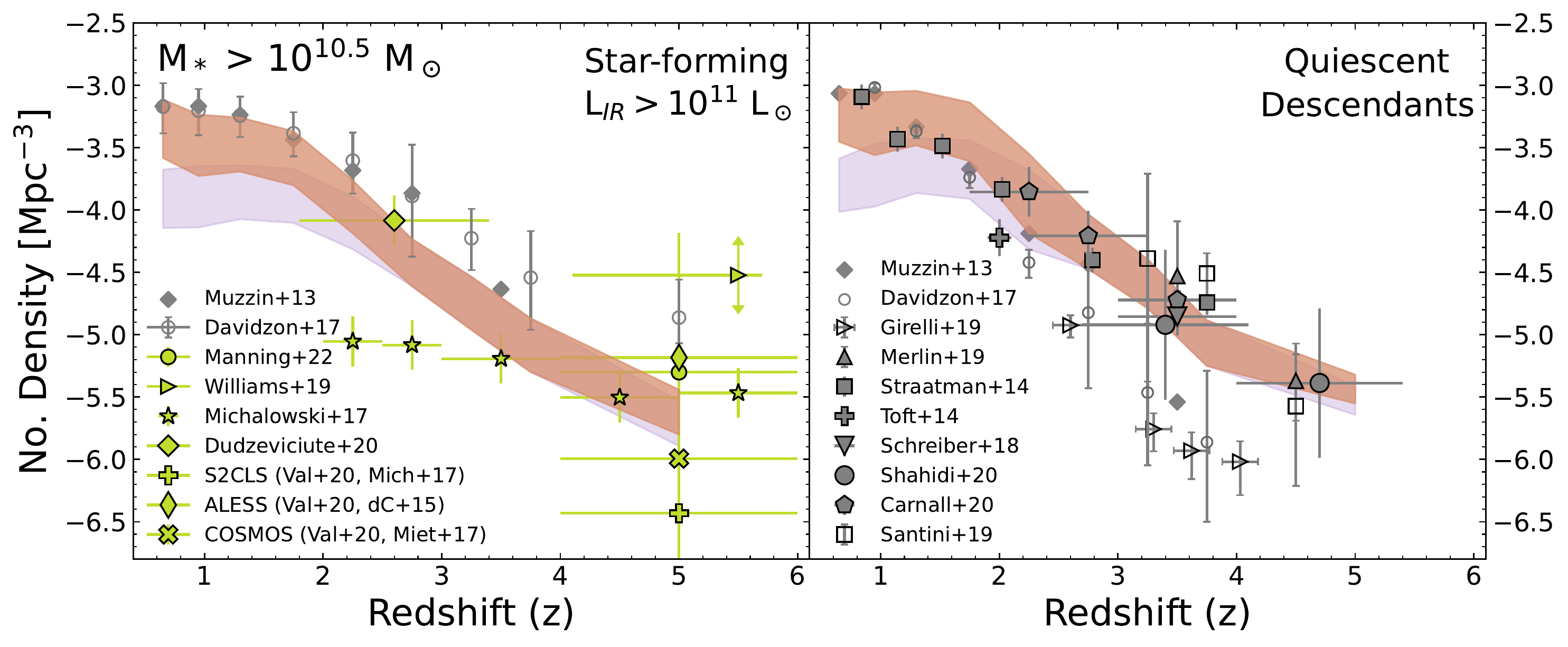}
\end{center}
\caption{The evolution of the number density of massive galaxies. Our model is presented as the pink band, with the non-duty-cycle corrected values represented by the purple band. \textbf{Left:} Observations of similar (but not necessarily equivalent, see Section \ref{sec:BigDOGnumbers}) DSFG populations are shown as green points \citep{michalowski17, Williams2019, dudz20,Valentino2020,Manning2022}, while UV/OP-based estimates are shown in grey \citep{muzzin13,davidzon17}. It is important to note that, for the observational data in green, the uncertainties on the abscissa demarcate the redshift bins of each observational sample. For example, when integrating this model over a similar redshift bin as the \citet{dudz20} sample, we achieve a similar number density estimate (demonstrated by the overlap between the $z\sim2$ portion of our model and the low redshift-end of the Dudzeviciute et al. sample). \textbf{Right:} Massive, passive galaxy number density constraints from the literature are shown as grey points \citep{muzzin13,Straatman2014,toft14,davidzon17,Schreiber2018,Santini2019,Girelli2019,Merlin2019,Shahidi2020,Carnall2020}. Note that the selection functions for the comparing quiescent galaxy samples vary significantly across these studies. Our results appear to align best with massive post-starburst (sSFR\,$<10^{-10}$/yr) galaxies. See Section \ref{sec:DescendantNumbers} for more discussion. 
} \label{fig:nodensities}
\end{figure*}

\section{Number and Stellar Mass Densities} \label{sec:inthesky}
For the remainder of this work, we focused on the massive, dust obscured population of galaxies predicted by our model, defined as all objects with M\,$\ge10^{10.5}$\,M$_\odot$ and L$_\mathrm{IR}\ge 10^{11}$\,L$_\odot$. We emphasize that the latter criterion emerged naturally in this model: nearly all ($\ge 99\%$) M\,$\ge10^{10.5}$\,M$_\odot$ star-forming galaxies produced by our model, across the redshift space probed here, have LIRG or greater IR luminosities; moreover, $>75\%$ of these massive star-forming galaxies have ultra-LIRG (ULIRG, L$_\mathrm{IR}\ge 10^{12}$\,L$_\odot$) luminosities or higher. We hereafter refer to these IR luminous massive galaxies as massive DSFGs.

\subsection{Duty Cycle Correction} \label{sec:dutycorr}
Gas depletion timescales measured in high-$z$ DSFGs suggest that the IR luminous phase of star formation is short lived \citep[on the order of tens to a few hundred Myr,][]{Tacconi2006, Barro2014, aravena16, Yang2017, D'Eugenio2020, Zavala2022}. This means that massive DSFGs are only traceable via their incredible IR emission for a small window of time, and therefore IR / sub-mm surveys are likely incomplete in capturing the DSFG population over cosmic timescales significantly wider than a few hundred Myr. If we assume that massive DSFGs are the primary progenitor population to massive quiescent galaxies at later times, we can derive a ``duty cycle correction factor'' that forces the two population densities to match over respective redshift windows \citep[as seen in e.g.][]{toft14, Valentino2020}. 

We iteratively tested a range of duty cycles from 50-1000\,Myr and found that a duty cycle correction factor of $\sim$\,500\,Myr was most effective in matching our predicted number densities to those of massive quiescent galaxies observed in the literature. Duty cycles on the order of a few hundred Myr or less result in a significant over production (10$\times$ more) of observed quiescent descendants across all epochs, while cycles closer to 1\,Gyr under produce the descendant population by 0.5\,dex at $z<2$. We note that a duty cycle correction of $\sim$\,500\,Myr only significantly impacts model results at $z<2$; all results at $z>2$ remain roughly the same in both the pre- and post-duty cycle corrected model as the widths of the redshift windows are approximately equivalent to the duty cycle correction factor. We show the pre- and post-duty cycle correction model predictions in Figure \ref{fig:nodensities} as purple and pink bands, respectively, and discuss further potential implications of this value in Section \ref{sec:duty_bursts}.

\subsection{Massive DSFG Number Densities} \label{sec:BigDOGnumbers}
In Figure \ref{fig:nodensities}, we show the redshift evolution of the number density of massive DSFGs (left), and of their quiescent descendants (right). During the $\sim1$\,Gyr between $z=3$ and 6, we predict an order of magnitude increase in number density of M\,$>10^{10.5}$\,M$_\odot$ dust obscured SF galaxies, similar to that seen in \citet{Marsan2022}. This growth rate tapers off at later times, taking another $\sim2$\,Gyr (from $z=3$ to 1.5) to achieve another order of magnitude of growth in population density. This $z>3$ evolution is steeper than that exhibited by UV/OP-based estimates from e.g. \citet{davidzon17}, where it takes an additional $\sim$\,half a billion years to achieve similar growth. In general, across $0.5<z<4$, UV/OP-bright massive SF galaxies are $\sim6-7\times$ more populous, except at $z=4-6$ where we find an order of magnitude more UV/OP-bright galaxies than massive DSFGs. (However, as shown in Section \ref{sec:dog_smf} and Figure \ref{fig:fractions}, the overall fraction of massive DSFGs compared to UV/OP-bright galaxies is larger at higher stellar mass thresholds.) 

One of the primary pieces of evidence that pins $z>3$ DSFGs as ancestors to giant ellipticals at later times is the similar co-moving number densities between the two populations \citep[e.g.][]{toft14}. However, the strength of this evolutionary picture is discrepant depending on the sample size, depths, and sensitivities of far-IR and sub-mm surveys used to characterized the massive, dust-obscured population. For example, \textit{Herschel} discovered DSFGs at $z>2$ are often comprised of a rare and extreme star-forming population (with SFRs up to and exceeding 1000\,M$_\odot$\,yr$^{-1}$); their number densities are insufficient to match those estimated for massive quiescent galaxies at $z>2$ \citep{Dowell2014, ivison2016, Duivenvoorden2018, maspitzer, Montana2021}. Instead, the recently discovered H-band dropouts, or alternatively the larger umbrella of galaxies that are optically ``invisible'' but sufficiently bright in the far-IR/sub-mm regime ($S_{850 \mu m} \gtrsim 1$\,mJy), may make up the bulk of the massive, star-forming galaxy population at $z>3$ \citep{Williams2019, Wang2019, Lagos2020, Manning2022}. Thus, it is perhaps more important to successfully model these more common, less extreme dusty, star-forming galaxies.

At $z=2-3$, when integrated over similar redshift ranges, our model successfully predicts similar space densities of the massive ($M_\star \gtrsim 3\times10^{10}$\,M$_\odot$) submillimeter galaxy populations presented in \citet[][$\sim 10^{-4}$\,Mpc$^3$ at $z=1.8-3.4$]{dudz20}, but predicts $5-10\times$ more massive DSFGs than what is seen in the \citet{michalowski17} SCUBA-2-detected sample in a similar redshift range. This difference is likely due to the employed SFR cutoff in the latter work, which only includes galaxies with SFRs $> 300$\,M$_\odot$\,yr$^{-1}$. In our model, $20-40$\% of M $\ge3\times10^{10}$\,M$_\odot$ galaxies at $z\sim2$ have SFRs $<300$\,M$_\odot$\,yr$^{-1}$ (with the percent decreasing with increasing redshift) -- this difference in population properties is sufficient to match the gap between our $z\sim2$ results and those seen in \citet{michalowski17}, within uncertainty.

At $z>3$, the model's success is difficult to quantify due to lack of strong observational constraints, including insufficient sample sizes of DSFGs and highly uncertain redshifts. Still, we find general agreement between the model and observations out to $z\sim5$, though observational uncertainties span two orders of magnitude. At $z=3-4$, the model predicts massive DSFG number densities in agreement (within uncertainty) with those of the \citet{michalowski17} sample (n\,$\sim1-2\times10^{-5}$\,Mpc$^3$). At $4<z<6$, we compare to the ALESS \citep{dacunha15}, COSMOS \citep{miett17}, and S2CLS \citep{michalowski17} samples as calculated in \citet[][pre-duty cycle correction]{Valentino2020}. To derive number densities for these surveys, \citet{Valentino2020} took strict photometric redshift constraints, requiring that both upper and lower limits on the photo-$z$s sit at or above $z=4$. As seen with the S2CLS derivations between \citet{michalowski17} and \citet{Valentino2020}, this requirement can reduce the number densities \textit{by an entire order of magnitude}, thereby demonstrating to first order one of the main difficulties in constraining DSFG number densities at $z>3$. Across these three surveys with conservative photo-$z$ restrictions, our model estimates lie between those derived from the COSMOS \citet{miett17} and ALESS \citet{dacunha15} samples, where n\,$\sim2\times10^{-6}\,$Mpc$^3$ at $4<z<6$, which is also roughly in alignment with \citet{michalowski17}.

Comparing to the observed number densities of the (potentially more populous) optically dark dust-obscured galaxies at $4<z<6$ is particularly difficult as observational constraints suffer either from small sample sizes (n\,$=1-4$, e.g. \citealt{Williams2019, Manning2022}), or are averaged over extremely large redshift ranges (e.g. $z=3-6$ in \citealt{Wang2019}). When comparing to estimates derived from a single exceptional object in \citealt{Williams2019}, our model predicts nearly 30$\times$ smaller populations of massive DSFGs. While the \citealt{Wang2019} sample of H-band dropouts includes a larger sample size ($n = 39$) over a larger survey area ($\sim$600\,arcmin$^2$), these objects span a (photometric redshift) range of $3<z<8$, with over half of the sample exhibiting redshift errors $\delta z \ge 1$. Still, when integrating our model across $3<z<5$, we find a similar space density as the authors (reported as n\,$\sim2\times10^{-5}$\,Mpc$^3$) as well as the entire sample of 2\,mm detected galaxies in \citet{Manning2022}. When including only galaxies with redshifts estimated at $4<z<6$, both the Wang et al. and Manning et al. estimates drop to n\,$\sim5-6\times10^{-6}$\,Mpc$^3$, which is closer to our model predictions. Unfortunately, due to the large uncertainties in the observational data, we limit and conclude our interpretation of the discrepancies between this work and the aforementioned optically-dim sources until further, more robust surveys are completed.

\begin{figure*}[ht]
\begin{center}
\includegraphics[trim=0cm 0cm 0cm 0cm, width=1.\textwidth]{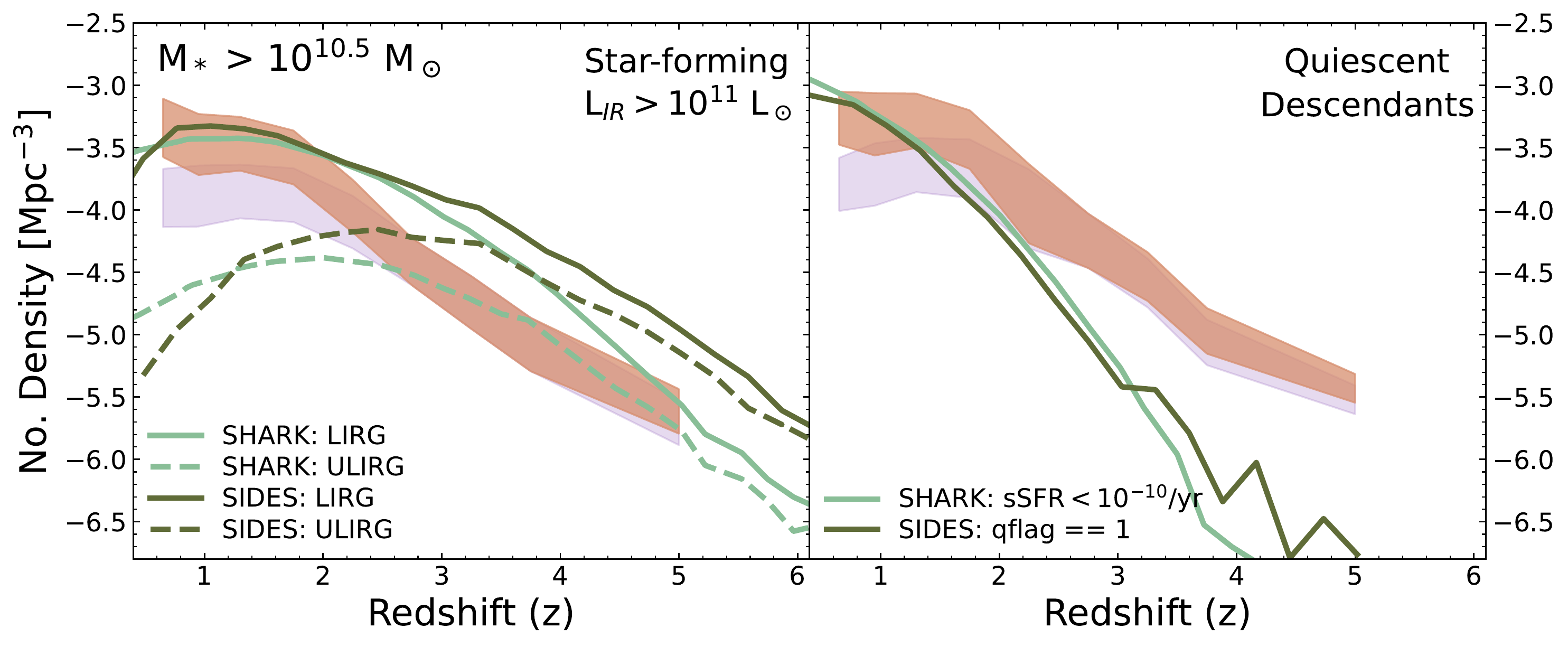}
\end{center}
\caption{The evolution of the number density of massive galaxies: our model versus other models. Our model is presented as the pink band, with the non-duty-cycle corrected values represented by the purple band. The SHARK model \citep{Lagos2018} is represented by the light green lines, while the SIDES model \citep{Bethermin2017} corresponds to the dark green lines. \textbf{Left:} Solid lines represent the massive LIRG (L$_\mathrm{IR} \ge 10^{11}$\,L$_\odot$) population, while dashed lines correspond to the the massive ULIRG (L$_\mathrm{IR} \ge 10^{12}$\,L$_\odot$). \textbf{Right:} The SHARK model quiescent galaxies are selected primarily as a post-starburst (sSFR\,$<10^{-10}$/yr) population as this is likely most in line with our mock descendant population, whereas the SIDES galaxies are selected from a publicly available catalogue and categorized as strictly passive. See Sections \ref{sec:BigDOGnumbers}-\ref{sec:nodensitymodels} for more details. 
} \label{fig:nodensities_models}
\end{figure*}

\subsection{Quiescent Descendants} \label{sec:DescendantNumbers}

\subsubsection{Generating Quiescent Descendants} \label{sec:q_smf}
The discovery and follow up observations of massive galaxies within the first 1-2\,Gyr of the Universe suggest that they form via extremely efficient and rapid star formation mechanisms during the early cosmos \citep[e.g.][]{Glazebrook2017, Schreiber2018, Schreiber2018b, Marsan2022, D'Eugenio2020, Valentino2020}. This is likely in the form of bursts of extreme star formation ($300-1000$\,M$_\odot$\,yr$^{-1}$) lasting $\sim100-300$\,Myr \citep[e.g.][]{Tacconi2006, Barro2014, aravena16, Yang2017, D'Eugenio2020, Zavala2022}, after which the cold gas reservoir is depleted, heated, and/or blown out. 

In our model, we make the following simplistic assumptions to derive descendant quiescent galaxy masses to compare with observations and models in the literature. We assume that the current SFR of a mock galaxy is maintained for an additional $50-300$\,Myr, in alignment with the gas depletion timescales seen in the aforementioned observational studies. These timescales are randomly sampled from a uniform grid with steps of $\delta_\mathrm{time} = 5$\,Myr. Then, we forward evolve the galaxy, adding the additional stellar mass built assuming a continuous SFR over the given timescale; to determine the redshift of the `quenched' galaxy, we add the same timescale to the galaxy's star-forming redshift (which was assigned in the initial model).

There are caveats to our simple forward evolution model: we do not include any prescriptions for AGN feedback, merging activity, additional gas accretion, or any other potential phenomenon that could theoretically slow down or speed up the SFRs / gas depletion timescales assigned to our mock galaxies. We see this as a strength of this model: there is no need to invoke complex assumptions on e.g. available gas mass reservoirs and potential heating mechanisms in order to successfully reproduce observed massive galaxy population densities (as shown in the proceeding Section). Furthermore, it is entirely possible that these more second-order corrections can be accounted for in the marginalization / uncertainty propagation process. For example, minor bumps or dips in the IRLF due to major mergers triggering IR-luminous star-formation would be accounted for in the uncertainty in the IRLF parameter space. Thus, while a more sophisticated model may be more ``complete'' in it's physical assumptions, it does not appear necessary to describe massive galaxy evolution to a first-order approximation.

\subsubsection{Quiescent Descendant Densities} 

On the right panel of Figure \ref{fig:nodensities}, we show the expected evolution in the number density of massive DSFG quiescent descendants derived in Section \ref{sec:q_smf}. In general, this model does an excellent job at modeling observed number counts in heterogeneous samples of massive, passive galaxies across $z=2-6$ \citep{muzzin13,toft14,Straatman2014,davidzon17,Schreiber2018,Merlin2019,Santini2019,Shahidi2020,Carnall2020}. This is particularly promising because of the heterogeneity across these samples: each of the aforementioned bodies of work has their own methods/thresholds of selection and characterization of passive galaxies, yet we are able to successfully reproduce similar estimates across such diversity. Some have slightly lower stellar mass thresholds \citep[e.g. M\,$>10^{10}$\,M$_\odot$,][]{Shahidi2020,Carnall2020}, or are only complete at higher stellar masses than those explored in this work \citep[e.g. M\,$>10^{11}$\,M$_\odot$ at $z=3-4$,][]{muzzin13}. Some employ specific SFR (sSFR\,$= $\,SFR/M$_\star$) cuts that capture post-starbursts \citep[e.g. log(sSFR/yr)\,$<-10$,][]{Straatman2014}, which are likely more in line with our model's assumptions / definition of massive DSFG descendants, while others focus strictly on fully quiescent galaxies \citep[e.g. log(sSFR/yr)\,$< -11$,][]{davidzon17}. 

Our model most disagrees with the work of \citealt{Girelli2019}, where we predict nearly an order of magnitude more passive galaxies at $3<z<4$. While Girelli et al. place similar stellar mass and sSFR cuts on their sample as some of the aforementioned studies, their quiescent galaxy selection technique is unique. They develop a color selection criteria based on mock quiescent galaxy colors generated from stellar population synthesis models. They employ an exponentially declining star formation history with a short $e$-folding time (100-300\,Myr) such that the galaxies are quiescent at the time of observation ($\sim$1-2\,Gyr into the Universe). In effect, this could mean that post-starburst galaxies are not captured in this color space as their colors are likely bluer than the objects discovered in Girelli et al. As mentioned before, it is likely that our model quiescent descendants fall in the post-starburst category, and therefore would be missed in the Girelli et al. study, possibly accounting for such a discrepancy in number densities. 

Like the ancestral massive DSFG population, this model predicts an order of magnitude increase in the number density of passive galaxies from $z=6$ to 3 (i.e. over $\sim1$\,Gyr), a growth rate similar to that seen in \citealt{Santini2019} and \citealt{Merlin2019}, and another order of magnitude over nearly twice the amount of time from $z=3$ to 1.5. This is unsurprising given the relatively short gas-to-stars duty cycles used in this work and exhibited by DSFGs at high redshifts \citep[e.g.][]{Tacconi2006, aravena16}.

\subsection{Comparison to Models}\label{sec:nodensitymodels}

In Figure \ref{fig:nodensities_models}, we compare our results to other models relevant to this work as evident by their established success in reproducing properties and/or observations of submillimeter-bright galaxies. First is the \textsc{Shark} semi-analytical model \citep{Lagos2018, Lagos2020}, which well-reproduced on-sky number counts at $\lambda=600\,\mu$m$-2$\,mm across $1<z<4$. \textsc{Shark} is a physical model that follows the formation of galaxies in a $\Lambda$CDM universe using a large suite of physical models for e.g. star formation, as well as stellar and AGN feedback. We also emply the Simulated Infrared Dusty Extragalactic Sky (SIDES) empirical model \citep{Bethermin2017}, which is successful in reproducing number on-sky counts of single-dish (i.e. confusion-limited) instruments, e.g. \textit{Herschel} PACS and SPIRE observations at $\lambda=70-500$\,$\mu$m. SIDES was built using compiled observational results on the SFMS and stellar-to-halo mass relationships. 

To compare the SIDES model, we use the publicly available catalogs,\footnote{SIDES catalogs can be found here: http://cesam.lam.fr/sides.} taking all galaxies with stellar mass $>10^{10.5}$\,M$_\odot$ in the same redshift bins as defined in Table \ref{table:n and smd}. We use the passive galaxies flag (\texttt{qflag == 1}) to separate star-forming and quiescent galaxies, and further divide the star-forming sample into LIRGs and ULIRGs (L$_\mathrm{IR}>10^{12}$\,L$_\odot$) by converting the provided SFRs into IR luminosities using the same conversion employed in our model (\citealt{Murphy2011,kennevans12}, see Section \ref{sec:IRLF}). To pull an analogous massive DSFG population from the SHARK model (C. Lagos, priv. communication), we use the same stellar mass, redshift, and IR luminosity criteria; for a comparative quiescent galaxy sample, we include post-starburst / near-quenching galaxies in the SHARK passive galaxy sample, defined as galaxies with log(sSFR/yr)\,$<-10$. Galaxy properties (i.e. SFR and stellar mass) in both populations were convolved with random Gaussian noise of width $\sim$\,0.2\,dex prior to calculating number and stellar mass densities.

At $z<2$, our (duty-corrected) model is in agreement with the massive LIRG (aka massive DSFG) estimates in both SHARK and SIDES. Beyond this redshift, the SIDES model predicts roughly three times as many massive DSFGs at all epochs, while the SHARK model predicts only twice as many up until $z\sim4$ when our results drop back into alignment. 

While \textit{all} mock massive DSFGs in this work have IR luminosities of LIRG status or brighter across $0.5<z<6$, the majority (75-99\%) of massive DSFGs exhibit even brighter ULIRG-like luminosities at $z\gtrsim2$. Thus, a more appropriate population comparison might be the ULIRG population in both the SIDES and SHARK models. Indeed, when comparing to the ULIRG population in both models, our results are significantly less discrepant, with a near-perfect agreement in predictions from the SHARK massive ULIRG model population at $z>2$. Consistency across the SHARK model and this work may be due to their careful and considerate treatment in relating dust surface densities with attenuation in the stellar continuum, including an energy balance in the far-infrared. One potential reason that the SIDES model, which is empirically-based and similar in build to this work, produces a higher density of massive IR luminous galaxies is due to differing star formation rate calculations. In SIDES, L$_\mathrm{IR}$ is converted to SFR using the \citealt{Kennicutt1998} equation, while in this work we use the relationship from \citealt{kennevans12}, which produces SFRs roughly 24\% lower than the former. We tested our model using the same L$_\mathrm{IR}$-to-SFR conversion as in SIDES and found that this SFR prescription brings our models into agreement for the massive ULIRG population, but our quiescent descendant population is overproduced by up to 0.5\,dex when compared to observations. 

As shown on the right side of Figure \ref{fig:nodensities_models}, both the SIDES and SHARK models produce significantly fewer massive quiescent galaxies at $z>2$ than this model. The SIDES team used SMF results from \citealt{davidzon17} to populate their model; this limits their model quiescent sample to strictly passive galaxies (log$_{10}$(sSFR/yr)\,$< -11$), while many of our model quiescent descendants are more likely post-starbursts (log$_{10}$(sSFR/yr)\,$< -10$). Thus, it is unsurprising that the SIDES quiescent number densities, like those reported in \citealt{davidzon17}, are up to a decade lesser in population density than our model predicts at $z>2$. For the SHARK model, we capture post-starburst \textit{and} strictly passive galaxies, yet see predictions as discrepant as those in the SIDES model. The massive end of the \textit{total} galaxy SMF in SHARK appears most sensitive to AGN feedback efficiency, but it is unclear if weaker AGN feedback would (or would not) produce sufficient populations of massive \textit{quiescent} galaxies to match our predictions at earlier times. 


\begin{deluxetable}{c cc | cc }[ht]
\centering
\tablecaption{Number densities and stellar mass densities as shown in Figures \ref{fig:nodensities} and \ref{fig:smds} for galaxies of stellar mass M\,$>10^{10.5}$\,M$_\odot$.}
 \tablehead{
 \multicolumn{1}{c}{ } & 
 \multicolumn{2}{c}{Massive DSFGs} & 
 \multicolumn{2}{c}{Quiescent Descendants}\\
 \colhead{Redshift} &  \colhead{log$_{10}$(n)} & \colhead{log$_{10}(\rho$)} & \colhead{log$_{10}$(n)} & \colhead{log$_{10}(\rho$)} }
 \startdata
 $0.5<z<0.8$ &  -3.37$^{+0.26}_{-0.21}$ & 7.23$^{+0.69}_{-0.42}$ & -3.23$^{+0.22}_{-0.21}$ &  7.28$^{+0.48}_{-0.36}$\\
 $0.8<z<1.1$ &  -3.49$^{+0.26}_{-0.23}$ & 7.25$^{+0.69}_{-0.44}$ & -3.32$^{+0.26}_{-0.24}$ &  7.35$^{+0.58}_{-0.39}$\\
 $1.1<z<1.5$ &  -3.48$^{+0.23}_{-0.20}$ & 7.30$^{+0.57}_{-0.39}$ & -3.26$^{+0.22}_{-0.21}$ &  7.45$^{+0.49}_{-0.36}$\\
 $1.5<z<2.0$ &  -3.60$^{+0.23}_{-0.20}$ & 7.29$^{+0.56}_{-0.38}$ & -3.37$^{+0.24}_{-0.23}$ &  7.44$^{+0.54}_{-0.38}$\\
 $2.0<z<2.5$ &  -3.97$^{+0.22}_{-0.21}$ & 7.13$^{+0.54}_{-0.41}$ & -3.84$^{+0.30}_{-0.34}$ &  7.31$^{+0.78}_{-0.55}$\\
 $2.5<z<3.0$ &  -4.42$^{+0.20}_{-0.18}$ & 6.81$^{+0.52}_{-0.37}$ & -4.26$^{+0.23}_{-0.21}$ &  6.95$^{+0.53}_{-0.38}$\\
 $3.0<z<3.5$ &  -4.75$^{+0.22}_{-0.20}$ & 6.58$^{+0.57}_{-0.45}$ & -4.60$^{+0.21}_{-0.18}$ &  6.64$^{+0.49}_{-0.36}$\\
 $3.5<z<4.0$ &  -5.07$^{+0.21}_{-0.22}$ & 6.26$^{+0.49}_{-0.43}$ & -5.07$^{+0.19}_{-0.17}$ &  6.27$^{+0.51}_{-0.38}$\\
 $4.0<z<6.0$ &  -5.62$^{+0.18}_{-0.18}$ & 5.53$^{+0.57}_{-0.41}$ & -5.43$^{+0.12}_{-0.11}$ &  5.95$^{+0.34}_{-0.31}$ \\
 \enddata 
\end{deluxetable} \label{table:n and smd}

\begin{figure*}[ht]
\begin{center}
\includegraphics[trim=0cm 0.8cm 0cm 0cm, width=1.\textwidth]{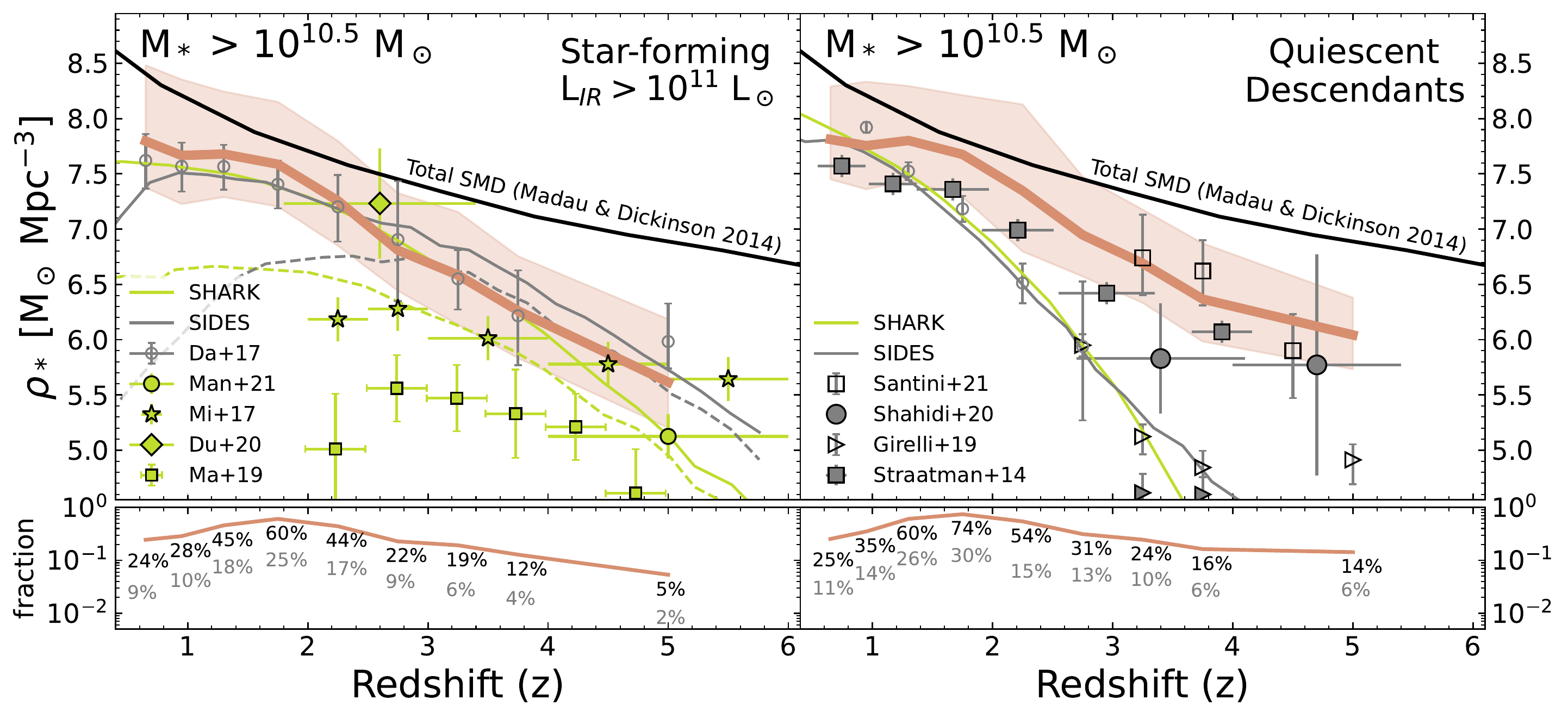}
\end{center} 
\caption{The evolution of the stellar mass density contributions by massive galaxies. Our model is represented by the pink band, with 1$\sigma$ uncertainties shown as the pale contours. The black line represents the integrated star formation rate density from \citet{madau14}. \textbf{Left:} Green points represent observations of massive DSFGs from the literature \citep{michalowski17, maspitzer,dudz20,Manning2022}; note that several of these are incomplete in stellar mass constraints for their DSFGs as the majority are insufficiently detected at rest-frame UV/optical wavelengths. Uncertainties on the abscissa represent redshift ranges. We also compare to UV/OP-bright galaxies derived in \citet{davidzon17} (grey points), and to massive LIRGs (solid) and ULIRGs (dashed) in the SHARK \citep[green][]{Lagos2020} and SIDES \citep[grey][]{Bethermin2017} models. \textbf{Right:} Grey points represent observations from the literature. SHARK (green) and SIDES model galaxies are selected in the same way as  in Figure \ref{fig:nodensities_models}. \textbf{Bottom:} The fraction of the total SMD (derived from the integrated SFRD) that our model massive DSFGs make up. The dark pink line and black percentages correspond to the pink band in the upper panels, while the light pink band and grey percentages represent the lower limits.
} \label{fig:smds}
\end{figure*}

\subsection{Stellar Mass Density}

Cosmic volume averaged studies over the last decade show how instrumental the dusty, star-forming galaxy population is in understanding stellar mass build up over cosmic time: their intense dust-obscured star formation rates dominate over unobscured (UV/OP-bright) star formation at $1 \lesssim z \lesssim 4$ \citep{madau14, casey2014dusty, Zavala2021}. However, the integrated cosmic star formation rate density (cSFRD) appears roughly 2$\times$ ($0.2-0.4$\,dex) greater than the measured cosmic stellar mass density out to $z\sim3$ \citep{madau14,Leja2015}, indicating an issue with `missing' stellar mass in direct measurements. Most of the potential explored drivers of this discrepancy are systemics such as overestimated star formation rates \citep{tomczak16}, underestimated stellar masses due to inflexible SED analysis \citep{Leja2015, Leja2020}, and photometry limitations \citep{madau14}. Missing from this conversation is a potential key component: it's the incomplete accounting of stellar mass from the massive dust-obscured population (see \citealt{maspitzer} for one of the only attempts). Owing to their heavily obscured nature, the field is still lacking in sufficient stellar continua observations for complete samples of DSFGs. One may argue that DSFGs are extremely rare and therefore not as significant in this discussion; however, as shown by the cSFRD, extreme properties of rare galaxies can be major -- or even dominant -- contributors when computing cosmic volume averaged quantities.

In Figure \ref{fig:smds}, we show the predicted contributions of massive DSFGs and their descendants to the cosmic stellar mass density (SMD). At $z=0.5$, this model predicts that massive DSFGs hold from one tenth to a quarter of the total SMD derived from the integrated cSFRD \citep{madau14}. This fraction increases to a peak around $z\sim2$, where massive DSFGs may contribute as much as 60\% of the SMD (and as little as 25\%). Such high fractions translate to a $0.1-0.3$\,dex increase on the observed SMD, which would be sufficient to bring the observed and integrated SMDs into alignment. Beyond $z\sim2$, the fraction steadily declines to $2-5\%$ at $z=4-6$, demonstrating how this population is likely less significant in cosmic volume studies in the high-$z$ Universe (possibly because this population may still be growing to sufficient sizes during these epochs). We see a similar rise and fall in the quiescent descendant population, reaching a maximum fraction of $30-74\%$ at $z\sim2$. 

When compared to other studies, we predict that the massive dust-obscured population is an equally important contributor to the SMD as the massive UV/OP-bright star forming galaxies captured in \citep{davidzon17}. Furthermore, this model is in rough agreement with the handful of existing constraints on observed massive DSFGs \citep[e.g.][]{michalowski17, dudz20, Manning2022}, with the exception of \citet{maspitzer} which probes a much more extreme and rare sub-population of massive DSFGs. We find that the lower SMD estimates of our quiescent descendant population is very similar to the \citealt{Straatman2014} ZFOURGE mass-limited (M\,$>10^{10.6}$\,M$_\odot$) sample, while our median prediction is in more agreement with \citealt{Santini2019}. Interestingly, our model results agree most with SMD predictions for the massive LIRG population in both the SIDES and SHARK models, whereas our predicted number density evolution was more in line with massive ULIRGs in both simulations. This likely means that the intrinsic stellar mass distributions in our model are skewed higher. In other words, the difference in SMDs is likely driven not by the number density evolution ($\Phi$) but by the shape of the high-mass end of the stellar mass function; our shallow broken power law extension beyond the knee likely yields a higher abundance of extremely massive galaxies. 

\section{Other Model Implications} \label{sec:otherimplications}

\subsection{AGN L$_{IR}$ Correction Is Necessary}
In Section \ref{sec:AGN}, we described how we corrected the mock galaxies' IR luminosities for AGN contamination so we could use L$_\mathrm{IR}$ to derive star formation rates, which were in turn used to derive stellar masses (via distributing SFRs over the star-forming main sequence). We tested the model results without the AGN correction procedure and found that the resulting SMF \textit{parameters} change by an average of $\sim5\%$. While seemingly insignificant, this propagates into an overestimate of the number density of massive DSFGs and their descendants by roughly half a dex at $z>2$. This is particularly relevant for studies with galaxy populations selected via IR emission at $\lambda = 8-50$\,$\mu$m wavelengths without sufficient multi-wavelength data to rule out major AGN contributions between $\lambda_\mathrm{rest} = 1-1000\,\mu$m. For these studies, it is possible that some of the selected galaxies have IR luminosities driven by AGN heating rather than by star-formation \citep{ciesla}. 


\subsection{The Duty Cycle \& Multiple Star Formation Bursts} \label{sec:duty_bursts}

In Section \ref{sec:q_smf}, we discuss how we create the mock quiescent descendant population by sampling gas depletion timescales of $50-300$\,Myr, and in Section \ref{sec:dutycorr}, we discuss how we correct the model number densities to match observed number counts using a duty cycle correction factor of 500\,Myr. Why don't these values match?

The range of gas depletion timescales used in this work are derived from direct measurements in the literature. They are generated using a \textit{closed-box} assumption such that given a galaxy's measured SFR and molecular gas mass, one can compute how long the galaxy could sustain such a star formation rate before depleting it's molecular gas reservoir. Yet, observations and simulations show that the star formation history of massive DSFGs is stochastic, with several bursts over short timescales due to a mix of mergers, sustained gas flows, and/or violent disk instabilities \citep{narayanan15, Tacchella2016, cowie18, Hodge2019, mcalpine, Lagos2020}. It can take up to 1\,Gyr for the halo to become hot enough (through e.g. AGN feedback or halo shock heating) such that galaxy-wide star formation finally abates. 

Thus, in the context of this model, we interpret these different time scales as referring to two distinct events. The first -- the gas depletion timescale -- refers to the \textit{current} burst of star formation, while the duty cycle correction refers to the overall ``average'' time it takes a massive DSFG to finally reach quiescence after several bursts.

\subsection{Relation to `Optically Dark' Galaxies}

One of the main goals of the current study was to predict the ``true'' massive DSFG number density at $z>0.5$ as the majority of DSFGs do not have sufficient rest-frame UV/optical detections to constrain their total stellar mass. A smaller portion of the massive DSFG population is near-invisible at these wavelengths and thus sometimes referred to as `optically dark' galaxies \citep{Wang2019, Williams2019, Talia2021, Manning2022}. This population is believed to be extremely dust-obscured, gas-rich, and already massive (M\,$\gtrsim10^{10.5}$\,M$_\odot$) at $z=3-6$. Furthermore, these studies posit that optically dark galaxies constitute the majority of the massive galaxy population in the first 2\,Gyr of the Universe, dominating in stellar mass density and in star formation rate density. 

Consistent with observations, our model also suggests a strong, if not dominating, prevalence of massive, heavily dust-obscured galaxies at similar epochs. It is interesting that we arrive at a similar conclusion using relatively modest assumptions on the properties and evolution of massive DSFGs. Moreover, our mock quiescent descendant population requires that some of these descendants are observed as high-$z$ post-starbursts (as opposed to strictly passive). This would manifest observationally as a significant population of massive gas-poor galaxies with young ages from recent bursts, yet strong Balmer breaks from evolved stellar populations that developed during earlier bursts. Indeed, recent studies on massive quiescent galaxies at $z\gtrsim3$ broadly corroborate these implications \citep{AlcaldePampliega2019, D'Eugenio2020, Forrest2020, Carnall2020}, suggesting a strong connection between massive DSFGs and quiescent galaxies in the early cosmos. Scheduled wide-field \textit{JWST} surveys with sufficient near-IR depths to capture these faint and rare populations such as the COSMOS-Web Survey (Casey et al., in prep) will be instrumental in constraining the stellar properties and number densities of these galaxy sub-types, and in more firmly investigating their connections through cosmic time.

\section{Conclusions} \label{sec:summary}

Owing to their extreme dust obscuration, much uncertainty still exists surrounding the stellar mass growth and content in dusty, star-forming galaxies at $z>1$. Observations and simulations establish strong links between DSFGs and massive quiescent galaxies at high-$z$, but a full characterization of the massive, dust-obscured galaxy population is still missing and critical to understanding massive galaxy evolution in the early cosmos. In this work, we present a novel numerical model that uses empirical data on dusty, star-forming galaxies to model their stellar mass contributions across the first $\sim$\,10\,Gyr of cosmic time. This method allows us to circumvent the observational limitations and restrictions that typically prevent galaxy evolution studies from fully capturing this extreme population. It also allows us to marginalize over the diverse range of properties observed in DSFG populations (e.g. position on the star-forming main sequence, gas depletion timescales), and to carefully fold in uncertainties in the underlying model assumptions (e.g. uncertainties in the parameters describing the IR luminosity function). Where possible, we take the most conservative and/or simple assumptions. 

With this model, we generate a dust-obscured stellar mass function with a high-mass power law extension that reaches beyond the `knee' (i.e. characteristic mass) of UV/OP-based SMFs in the literature. We find that two SMF parameters -- the number density evolution ($\Phi_\star$) and the slope of the high mass end ($\beta$) -- are not significantly dependent on whether more DSFGs are starbursts (and therefore lower in stellar mass) or main sequence galaxies (and therefore higher in stellar mass). Almost all ($\sim99\%$) of the galaxies at massive-end of our dust-obscured SMF (at M\,$\ge10^{10.5}$\,M$_\odot$) have LIRG-like luminosities (L$_\mathrm{IR} \ge 10^{11}$\,L$_\odot$), with the majority ($>75\%$) exhibiting ULIRG luminosities or greater (L$_\mathrm{IR} \ge 10^{12}$\,L$_\odot$). When compared to UV/OP-based estimates, we predict massive DSFGs are equally if not more prevalent in number at the massive end of galaxy evolution at $z>1$, constituting as much as $50-100\%$ of all massive star-forming galaxies at  M\,$\ge10^{11}$\,M$_\odot$.

We predict the volume averaged number density of massive DSFGs and find general agreement with observations, but the lack of sufficient measurements on real, observed DSFG stellar masses leave up to two orders of magnitude in uncertainty at $z>4$. What is perhaps more striking is that, after forward modeling mock massive DSFGs to their quiescent descendants, we find remarkable agreement with our predictions and the number densities of observed massive passive galaxies at high-$z$. This means that, to first order, massive dust-obscured galaxies with ULIRG-like luminosities or greater are a sufficient ancestral population to describe the prevalence of massive quiescent galaxies at $z>1$. Between the two populations (massive DSFGs and their quiescent descendants), we predict an intense epoch of growth during the $\sim1$\,Gyr from $z=6$ to 3 during which the majority of the most massive galaxies at high-$z$ grow and then quench. We also show that massive DSFGs and their quiescent descendants are significant contributors to the cosmic stellar mass density, constituting as much as $25-60\%$ during the peak of cosmic star formation at $z\sim2$; this could be sufficient in closing the gap between the integrated cosmic star formation rate density and the directly measured stellar mass density.

In order to close the uncertainty on this model's implications and better understand massive galaxy evolution in the first 2\,Gyr of the cosmos, future studies should focus on collating large samples of DSFGs (perhaps complete down to ULIRG-like luminosities) with both strong redshift constraints and deep near-IR coverage. What drives the era of massive galaxy growth between $z=3-6$? What astrophysical processes precipitate extreme rates of star-formation (and dust production) so early on? And what causes such massive galaxies to rapidly quench during an epoch where cold molecular gas is so readily available for most other galaxies? We can only hope to answer these questions through a coordinated symbiosis between observatories across the electromagnetic spectrum. Specifically, we will be able to test, explore, and constrain the implications of this work by coupling deep near-IR observations afforded by \textit{JWST} with sensitive, high-resolution submillimeter interferometers such as ALMA. 

\acknowledgments

ASL acknowledges support for this work provided by NASA through the NASA Hubble Fellowship Program grant \#HST-HF2-51511.001-A, awarded by the Space Telescope Science Institute, which is operated by the Association of Universities for Research in Astronomy, Inc., for NASA, under contract NAS5-26555. ASL also acknowledges support from the Ford Foundation. CMC thanks the National Science Foundation for support through grants AST-1814034 and AST-2009577 as well as the University of Texas at Austin College of Natural Sciences for support; CMC also acknowledges support from the Research Corporation for Science Advancement from a 2019 Cottrell Scholar Award sponsored by IF/THEN, an initiative of Lyda Hill Philanthropies. This research made use of Astropy,\footnote{http://www.astropy.org} a community-developed core Python package for Astronomy \citep{astropy:2013, astropy:2018, astropy:2022}.

\pagebreak

\bibliographystyle{aasjournal}
\bibliography{references}

\end{document}